\documentstyle[aps,epsf,graphicx,eqsecnum,preprint]{revtex}
\newcommand{\be}{\begin{equation}}
\newcommand{\ee}{\end{equation}}
\newcommand{\bea}{\begin{eqnarray}}
\newcommand{\eea}{\end{eqnarray}}
\def \in{\leftskip = 40 pt\rightskip = 40pt}

\def \out{\leftskip = 0 pt\rightskip = 0pt}
\def\nn{\nonumber\\}

\def\frak#1#2{{\textstyle{{#1}\over{#2}}}}

\def\pa{\partial}
\def\semi{;\hfil\break}

\def\Ytilde{\tilde Y}
\def\sy{supersymmetry}   
\def\sic{supersymmetric} 
\def\DRED{\ifmmode{{\rm DRED}} \else{{DRED}} \fi}
\def\DREDp{\ifmmode{{\rm DRED}'} \else{${\rm DRED}'$} \fi}  
\def\NSVZ{\ifmmode{{\rm NSVZ}} \else{{NSVZ}} \fi}

\def\npb{{Nucl.\ Phys.\ }{\bf B}}
\def\prd{{Phys.\ Rev.\ }{\bf D}}
\def\prl{Phys.\ Rev.\ Lett.\ }
\def\plb{{Phys.\ Lett.\ }{\bf B}}

\def\Ycal{{\cal Y}}
\def\mtilde{\tilde m}
\def\thbar{\bar\theta}
\def\Phbar{\bar\Phi}
\def\Dbar{{\bar D}}
\def\mbar{{\bar m}}

\def\bxhat{\hat\beta_{\xi}}
\def\lf{16\pi^2}
\def\llf{(16\pi^2)^2}
\def\lllf{(16\pi^2)^3}
\def\Tr{\hbox{Tr}}
\def\pslash{p\!\!\! /}
\def\kslash{k\!\!\! /}
\def\qslash{q\!\!\! /}
\def\adot{\dot \alpha}
\begin{document}

\begin{titlepage}
\begin{flushright}
LTH 480\\
hep-ph/0007291\\
\end{flushright}

\vspace*{3mm}

\begin{center}
{\Huge
The Fayet-Iliopoulos $D$-term and its renormalisation in softly-broken
supersymmetric theories}\\[12mm]
 
{\bf I.~Jack, D.R.T.~Jones and S.~Parsons} \\
   
%\end{center}   
 
\vspace{8mm} 

Dept. of Mathematical Sciences,
University of Liverpool, Liverpool L69 3BX, UK\\
\end{center}

\vspace{3mm}
\begin{abstract}

We consider the renormalisation of the  Fayet-Iliopoulos $D$-term in a
softly-broken  abelian supersymmetric theory, and calculate  the
associated $\beta$-function through three loops.  We show that there
exists (at least through three loops)  a renormalisation group invariant
 trajectory for the coefficient of the $D$-term,  corresponding to the
conformal anomaly solution for the soft masses and couplings. 
 
\end{abstract}

\end{titlepage}

\section{Introduction}

In abelian gauge theories with $N=1$ supersymmetry 
there exists a possible 
invariant that is not allowed in the non-abelian case: the 
Fayet-Iliopoulos $D$-term,
\be
L =
\xi\int V (x,\theta, \thbar)\,d^4\theta  = \xi D(x).
\label{dta}
\ee
In this paper we discuss the 
renormalisation of $\xi$ in the presence of 
the standard soft supersymmetry-breaking terms 
\be
L_{\rm SB}=(m^2)^j_i\phi^{i}\phi_j+
\left(\frac{1}{6}h^{ijk}\phi_i\phi_j\phi_k+\frac{1}{2}b^{ij}\phi_i\phi_j
+ \frac{1}{2}M\lambda\lambda+{\rm h.c.}\right)
\label{Aaf}
\ee

Let us begin by reviewing the position when there is no
supersymmetry-breaking,  i.e. for $L_{\rm SB}=0$.  Many years ago,
Fischler et al\cite{fisch} proved
an important  result concerning  the renormalisation of the $D$-term
(see also  Ref.~\cite{shwg}).   Since it is a
$\int\,d^4\theta$-type term, one may expect that the $D$-term  will
undergo renormalisation in general. Moreover, by simple power-counting 
it is easy to show that the said renormalisation is in  general
quadratically divergent. Evidently this poses a naturalness  problem
since (if present) it would introduce the cut-off mass scale  into the
scalar potential. At the one loop level it is easy  to show that the
simple condition $\Tr \Ycal = 0$ (where $\Ycal$ is  the $U_1$
hypercharge and the trace is taken over the chiral supermultiplets) 
removes the divergence. Remarkably, although one may of course  easily
draw individual diagrams proportional (for example)  to $\Tr \Ycal^5,
\Ycal^7 \cdots$ etc.,  this condition suffices to all orders.

In the presence of supersymmetry breaking, however, it is clear that $\xi$ will 
suffer {\it logarithmic\/} divergences. If calculations are done  in the
component formalism with $D$ eliminated by means of its equation of 
motion, then these divergences are manifested via contributions to  the
$\beta$-function for $m^2$. It is in this
manner that the results for the soft $\beta$-functions were given 
in, for example,  Ref.~\cite{jj}.  
Here we prefer to consider the renormalisation of $\xi$
separately; an  advantage of this is that it means that the exact
results for the soft $\beta$-functions presented in 
Refs.~\cite{jjf}-\cite{jjpa} 
(see also Refs.~\cite{yam}\cite{ahetal})  
apply
without change to the abelian case.
The result for $\beta_{\xi}$ is as follows:
\be
\beta_{\xi} = \frac{\beta_g}{g}\xi + \bxhat
\label{exacta}
\ee
where $\bxhat$ is determined by $V$-tadpole (or in components 
$D$-tadpole) graphs, and is independent of $\xi$. In the 
supersymmetric case, we have $\bxhat = 0$, whereupon Eq.~(\ref{exacta}) 
is equivalent 
to the statement that the $D$-term, Eq.~(\ref{dta}), is unrenormalised. 
In the presence of Eq.~(\ref{Aaf}), however, $\bxhat$ depends on 
$m^2$, $h$ and $M$ (it is easy to see that it cannot depend on $b$). 
The main result of this paper is a complete calculation of 
$\bxhat$ through three loops; 
it is interesting that the dependence on  $h$ and $M$ arises 
first at this order. (A partial calculation was 
presented in Ref.~ \cite{jjdterm}.)

Although in this paper we restrict ourselves to the abelian case, 
it is evident that a $D$-term can occur with a direct product gauge group
($G_1\otimes G_2\cdots$) if there is an  abelian factor: 
as is the case for the  MSSM. In the MSSM context one may 
treat $\xi$ as a free parameter at the weak scale\cite{gouv}, 
in which case there is no need to know $\bxhat$. However, if 
we know $\xi$ at gauge unification, then we need $\bxhat$ to predict 
$\xi$ at low energies.  Now in the $D$-uneliminated case it is possible 
to express all the $\beta$-functions associated with the soft 
supersymmetry-breaking terms given in Eq.~(\ref{Aaf}) in terms of the gauge 
$\beta$-function $\beta_g$, the chiral supermultiplet anomalous 
dimension $\gamma$ and a certain function $X$ which appears only 
in $\beta_{m^2}$; moreover in a special renormalisation 
scheme (the NSVZ scheme), $\beta_g$ can also be 
expressed in terms of $\gamma$, and $X$ takes a particularly simple form
\cite{jjpa}\cite{kkz}. 
It is clearly of interest to ask whether an analogous exact expression
exists for $\beta_{\xi}$. Moreover, there exists an exact solution 
to the soft RG equations for $m^2$, $M$ and $h$ corresponding to 
the case when all the supersymmetry-breaking arises from the 
conformal anomaly\cite{amsbrefs}
and it is also interesting to ask whether this solution can be 
extended to the non-zero $\xi$ case.

The key to the derivation of the exact results for the soft $\beta$-functions
is the spurion formalism. The obstacle to deriving an analogous result for 
$\beta_{\xi}$ is the fact that individual superspace diagrams are (as already 
mentioned) quadratically divergent. We do, however, present 
a solution for $\xi$ related to the conformal anomaly solution, but which 
must be constructed order by order in perturbation theory.     

\section{Renormalisation and non-propagating fields}
\label{sec:nonpop}
\subsection{The non-\sic\ case}

This paper is concerned with the renormalisation of the coefficient 
of an auxiliary field term, and it is perhaps useful to begin with a 
(we hope) pedagogical discussion of this in a non-supersymmetric context.
One often sees the statement that the field theory
\be
L = \frac{1}{2} (\pa_{\mu}\phi)^2 + \frac{1}{2}F^2 +\frac{1}{2}hF\phi^2
\label{Aa}
\ee
where $\phi^2 = \sum_{a = 1}^{a= N} \phi^a\phi^a$,  
is equivalent to the theory
\be
L = \frac{1}{2} (\pa_{\mu}\phi)^2  - \frac{\lambda'}{24}\phi^4
\label{Aaa}
\ee
(where $\phi^4 = (\phi^2)^2$), 
by virtue of the equation of motion for the non-propagating field $F$,
which is 
\be
F = - \frac{1}{2}h\phi^2
\ee
so that 
\be
\lambda' = 3h^2.
\label{Aab}
\ee
There is a trap for the unwary here, however, in that Eq.\ (\ref{Aa}) 
is not multiplicatively 
renormalisable, and as a consequence Eq.\ (\ref{Aab}) is not 
renormalisation group (RG) invariant. Let us replace Eq.\ (\ref{Aa}) by 
\be
L = \frac{1}{2} (\pa_{\mu}\phi)^2 + \frac{1}{2}F^2 +\frac{1}{2}hF\phi^2
-\frac{\lambda}{24}\phi^4.\label{Ab}
\ee
We then obtain (eliminating $F$) 
\be
\lambda' = \lambda + 3h^2,
\label{Ac}
\ee
and it follows that
\be
\beta_{\lambda'} = \beta_{\lambda} + 6h\beta_h,
\label{Ad}
\ee
which is easy to verify at one loop by direct calculation:

\begin{mathletters}
\label{Mmfij}
\bea
\lf\beta_{\lambda'}& = &\frac{N+8}{3}{\lambda'}^2\label{Mmfij:1} \\
\lf\beta_{\lambda}& = &\frac{N+8}{3}\lambda^{2} +12\lambda h^2 + 12h^4
\label{Mmfij:2} \\
\lf\beta_h & = & \frac{N+4}{2}h^3 + \frac{N+2}{3}h\lambda
\label{Mmfij:3} 
\eea
\end{mathletters}
and it is easy to see that  Eq.\ (\ref{Ad}) indeed 
holds. The minor subtlety here 
is that $\beta_{\lambda}$ does not vanish when $\lambda =0$, so that 
the naive relation Eq.\ (\ref{Aab}) is {\it not\/} RG invariant. 
Consequently, if we set $\lambda = 0$, then Eqs.\ (\ref{Mmfij:1}) 
and (\ref{Mmfij:3}) and are not 
compatible with the (naive) result of taking $\mu\frac{d}{d\mu}$ of 
Eq.\ (\ref{Aab}). 

One may generalise this example as follows, by introducing a mass for $\phi$ 
and a linear $F$-term:
\bea
L & = &\frac{1}{2} (\pa_{\mu}\phi)^2 -\frac{1}{2}m^{2}\phi^2 
+ \frac{1}{2}F^2 +\xi F +\frac{1}{2}hF\phi^2-\frac{\lambda}{24}\phi^4\nn
& = & \frac{1}{2} (\pa_{\mu}\phi)^2 -\frac{1}{2}{m'}^2\phi^2
-\frac{\lambda'}{24}\phi^4
\label{Af}   
\eea
where 
\be
{m'}^2 = m^{2} + h\xi.
\ee
We now have the additional identity
\be
\beta_{{m'}^2} = \beta_{m^{2}} + h\beta_{\xi} + \xi\beta_h
\ee
which can be verified at one loop using the results:
\bea
\lf\beta_{{m'}^2} & = & \frac{N+2}{3}\lambda'{m'}^2\nn
\lf\beta_{\xi} & = & \frac{N}{2}h^2\xi + Nhm^{2}\nn
\lf\beta_{m^{2}} & = & \frac{N+2}{3}\lambda m^{2} + 2h^2m^{2}
\eea
together with the result for $\beta_h$ which is unaffected.

\subsection{The supersymmetric case: $D$-terms}

After this warm-up exercise, let us turn to a softly-broken 
abelian  \sic\ gauge theory.
The relevant part of the Lagrangian is as follows:
\be
L = \frac{1}{2}D^2 +  \xi D + gD\phi^*\Ycal\phi - \phi^* m^2\phi+\cdots 
\label{sscone}\ee
where  $\Ycal^i{}_j$ is the charge matrix of the chiral supermultiplet, 
and $m^2$ is a supersymmetry-breaking term. 
After eliminating $D$ this becomes
\be
L = -\phi^* \mbar^2\phi - \frac{1}{2}g^2(\phi^*\Ycal\phi)^2,
\ee
where
\be
 \mbar^2 = m^2 + g\xi \Ycal.
\label{lsusyb}\ee
RG invariance of this result gives  
\bea
\beta_{\mbar^2}(\mbar^2, \cdots) & = & \beta_{m^2}(m^2, \cdots) 
+ \beta_g\xi \Ycal + g \beta_{\xi}\Ycal\nn
& = & \beta_{m^2}(m^2, \cdots) 
+ 2\beta_g\xi \Ycal + g \Ycal \bxhat(m^2,\cdots),
\label{appe}
\eea
where
\be
\beta_{\xi} = \frac{\beta_g}{g}\xi + \bxhat
\label{appf}
\ee
with $\bxhat$ independent of $\xi$. For a derivation of 
Eq.\ (\ref{appf}), see Ref.~\cite{jjdterm}.
What about the pitfall in the toy model which led us to introduce 
$\lambda$? We are saved by \sy: if we add a $\phi^4$ term to 
Eq.\ (\ref{sscone}), then 
\sy\ would be broken, at the dimension 4 level; 
contrariwise, if we omit it then it will 
not be generated. Therefore, Eq.\ (\ref{appe}) is valid.

There is an important distinction between 
$\beta_{\mbar^2}(\mbar^2, \cdots)$ and $\beta_{m^2}(m^2, \cdots)$, which 
both appear in Eq.\ (\ref{appe}), and determine the mass 
renormalisation with $D$ eliminated and uneliminated respectively. 
 Because $\beta$-functions are determined 
by 1PI diagrams, $\beta_{m^2}$ does not contain any $D$-tadpole contributions;
the renormalisation of these is dealt with separately by $\beta_{\xi}$. 
However, in the $D$-eliminated formalism, there is no $\beta_{\xi}$, 
and there is a distinct set of contributions to $\beta_{\mbar^2}$ 
involving the  four-point vertex created by eliminating $D$. 
It follows that 
\be
\beta_{\mbar^2}(\mbar^2,\cdots) = \beta_{m^2} (\mbar^2,\cdots) 
+ g\Ycal \bxhat(\mbar^2, \cdots),
\label{appi}
\ee
since diagrams corresponding to one or more insertions 
of a $D$-tadpole type contribution on the {\it internal\/} line of a 
diagram do not contribute to the $\beta$-function because 
the corresponding Feynman integral is factorised~\cite{jjold}.

Let us now define our notation for the calculation.
We take an abelian  $N=1$ supersymmetric gauge theory with
superpotential
\be
W(\Phi)=\frac{1}{6}Y^{ijk}\Phi_i\Phi_j\Phi_k+   
\frac{1}{2}\mu^{ij}\Phi_i\Phi_j,
\label{eqf}
\ee
and at one loop we have
\begin{mathletters}
\bea
\label{beone}
\lf\beta_g^{(1)} &=& g^3Q = 
g^3\Tr\left[\Ycal^2\right], \label{beone:1}\\
\lf\gamma^{(1)i}{}_j &=& P^i{}_j 
=\frac{1}{2}Y^{ikl}Y_{jkl}-2g^2(\Ycal^2)^i{}_j. 
\label{beone:2}
\eea
\end{mathletters}

In the spurion formalism the soft-breaking Lagrangian is given by
\bea
L_{\rm soft}&=&\int d^2\theta \theta^2\left(\frac{1}{6}h^{ijk}\Phi_i\Phi_j\Phi_k
+\frac{1}{2}b^{ij}\Phi_i\Phi_j+\frac{1}{2}MW^{\alpha}W_{\alpha}\right) 
+\hbox{h.c.}\\
&&+\int d^4\theta (m^2)^i{}_j\theta^2\thbar^2\Phbar_ie^{-g\Ycal V}\Phi^j,
\label{softsup}
\eea
where $V$ is the vector superfield and $W^{\alpha}$ 
the corresponding field strength.
The equivalent expression in terms of components is given in 
Eq.\ (\ref{Aaf}).
With the explicit all orders result for $\beta_{m^2}$, we 
prove a remarkably simple result for $\bxhat$.
The afore-mentioned exact result for $\beta_{m^2}$ is~\cite{jjp} 
\be
(\beta_{m^2})^i{}_j (m^2, \cdots) =\left[
2{\cal O}{\cal O}^* +2MM^{*} g^2{\pa
\over{\pa g^2}} +\Ytilde{\pa\over{\pa Y}}
+\Ytilde^{*}{\pa\over{\pa Y^{*}}} +
X\frac{\pa}{\pa g}\right]
\gamma^i{}_j.
\label{Ajy}
\ee
where
\be
{\cal O}=\left(Mg^2{\pa\over{\pa g^2}}-h^{lmn}{\pa
\over{\pa Y^{lmn}}}\right),
\label{Ajb}
\ee
\be
\Ytilde^{ijk}=(m^2)^i{}_lY^{ljk}+(m^2)^j{}_lY^{ilk}+(m^2)^k{}_lY^{ijl}
\label{Ajd}
\ee
and (in the NSVZ scheme) 
\be
\lf X^{\NSVZ}=-2g^3 \Tr \left[m^2\Ycal^2\right].
\label{exX}
\ee
Once again we should emphasise that, whereas in a non-abelian theory 
Eq.\ (\ref{Ajy})\ holds in both the $D$-eliminated and $D$-uneliminated 
formalism, in a theory with abelian factors it is only true 
for $D$ uneliminated. 

It is now easy to show that 
\be
\beta_{m^2} (\mbar^2,\cdots) = \beta_{m^2} (m^2,\cdots).
\label{appj}
\ee
This follows simply by substituting for $\mbar^2$ from Eq.~(\ref{lsusyb}) and 
then using the facts that 
\be
(\Ycal)^i{}_lY^{ljk}+(\Ycal)^j{}_lY^{ilk}+(\Ycal)^k{}_lY^{ijl} = 0
\label{appk}
\ee
by gauge invariance, and 
\be
\Tr (\Ycal^3) = 0
\label{appl}
\ee
for anomaly cancellation. 

The result for $X$, 
Eq.~(\ref{exX}), applies in the NSVZ scheme, which is one of a class of schemes 
related by redefinitions of $g$ and $M$, the ramifications of which 
are described in Ref.~\cite{jjf}. Now $X$ transforms non-trivially 
under these redefinitions\cite{jjpa}, 
but it can be shown using Eqs.~(\ref{appk}), 
(\ref{appl}) that $X$ is unchanged by the replacement 
$m^2\to \mbar^2$ in any member of this class of schemes; 
consequently Eq.~(\ref{appj}) always applies.
We then find immediately from Eqs.~(\ref{appe}), (\ref{appi}) that:
\be
\bxhat(\mbar^2,\cdots) = 2\frac{\beta_g}{g}\xi + \bxhat(m^2,\cdots).
\label{appx}
\ee

Now on dimensional grounds we may write: 
\be
\bxhat = m^2 A_1(g, Y, Y^*) + hh^* A_2 (g, Y, Y^*)
+MM^* A_3(g, Y, Y^*) + (Mh^* + M^*h)A_4 (g, Y, Y^*),
\label{apph}
\ee
where we have suppressed $(i,j\cdots)$ indices for simplicity.
(In the conventional DRED 
scheme, $\bxhat$ will also depend on the $\epsilon$-scalar $(\hbox{mass})^2$, 
$\mtilde^2$, and this dependence, as we shall see, arises first at three 
loops. Our three-loop result, therefore, will be in the $\DREDp$ scheme
\cite{jjmvy}.)
Hence we have at once that
\be
\Tr(\Ycal A_1) = 2\frac{\beta_g}{g^2}.
\label{appm}
\ee
So if we take the $D$-tadpole contributions to $\beta_{\xi}$, then 
the terms proportional to $m^2$ will reduce to $2 \beta_g/g$ if we replace 
$m^2$ by $g\Ycal$. This result is, in fact, clear from a diagrammatic 
point of view, since the aforesaid replacement converts the diagrams 
into $D$ self-energy graphs, and hence indeed gives rise to $\beta_g$.

\section{The one loop calculation}

Here we describe the one-loop calculation of $\bxhat$; this is
straightforward,  of course. In a softly 
broken theory, the $\bxhat$ calculation  may be carried out in components, or
using the superfield spurion formalism. Usually, superfield 
techniques (once mastered) offer a substantial reduction in labour 
compared to component calculations; we will begin to suspect 
from the one loop calculation, however, and confirm in the next section, 
that this is not the case here.  

\bigskip
%\begin{figure} 
%\caption{\label{figone}}
\epsfysize= 1.0in
\centerline{\epsfbox{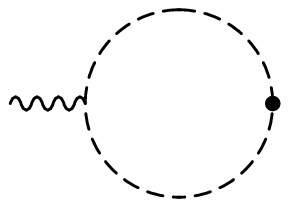}}
\in
{\it \noindent Fig.~1:
Feynman diagram for the one-loop calculation in components.
Dashed lines are scalar propagators and the
external field is a $D$. Blobs denote $m^2$ insertions.
}
\bigskip
\out
%\end{figure}

In components there is a single diagram, shown in Fig.~1, and we have
\bea
\hbox{Fig.~1} & = &  -g(\Ycal)^i{}_j\int d^d k
\left(\frac{1}{k^2 + m^2}\right)^j_i\nn
 & = &  -g(\Ycal)^i{}_j\int d^d k\left(\frac{1}{k^2}\delta^j{}_i
 - \frac{1}{k^4}(m^2)^j{}_i+\cdots \right)\nn 
& = & -g\Tr(\Ycal)\int \frac {d^d k}{k^2} +
g\Tr(\Ycal m^2)\frac{2}{\lf(4-d)} + \hbox{finite terms,}
\label{compone}
\eea
where our integration measure $d^dk$ includes the usual $(2\pi)^{-d}$ factor. 
In order to extract the ultra-violet divergence from the 
logarithmically divergent term in Eq.~(\ref{compone})\ we 
have made the replacement
\be
\int\frac{ d^d k }{k^4}\to \int\frac{d^d k}{(k^2 + m_0^2)^2}
\ee
where $m_0$ is an infra-red (IR) regulator mass. Naturally we 
could have directly evaluated the diagram without first 
expanding in powers of $m^2$, but this procedure would be 
cumbersome at higher loops; it is simpler to treat $m^2$ as 
an insertion and introduce regulator masses only for 
those propagators which are IR-dangerous. This technique was described
in Ref.~\cite{vlad}, and is generally more convenient than the 
alternative of ``threading'' a single momentum through the 
diagram\cite{jones}.  The pole result for 
a graph of any number of loops, {\it when all sub-divergences are 
subtracted}, is independent of the precise details of how the IR 
divergences are regulated. All this, is of course, well known 
to higher-loop calculators but may, perhaps 
be of some pedagogical interest. 

We see that to remove the quadratic divergence we must have $\Tr \Ycal = 0$, 
and that at one loop 
\be
\bxhat = \frac{1}{\lf}g\Tr(\Ycal m^2).
\ee
In the superfield spurion calculation we have two graphs, shown in Fig.~2.

\bigskip
\epsfysize= 1.0in
\centerline{\epsfbox{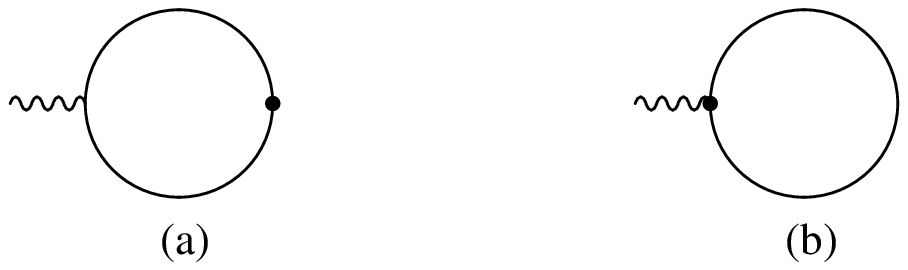}}
\in
{\it \noindent Fig.~2:
Feynman diagrams for the one-loop calculation in superspace.
Solid lines are chiral propagators and the external lines are vector 
superfields. Blobs denote $m^2$ insertions.}
\bigskip
\out

The results are as follows:
\bea 
\hbox{Fig.~2a} & = & 
-g\int d^d k \int d^4\theta\, V(\theta,\thbar) D^2\Bigl[
\Tr(\Ycal)\frac{\delta^{(4)} (\theta - \theta')}{k^2}\nn & + &
\Tr\left(\Ycal m^2\right)\frac{1}{k^4}\Dbar^2\theta^2\thbar^2 D^2
\delta^{(4)} (\theta - \theta')\Bigr]\Dbar^2 |_{\theta = \theta'}
\label{oness}
\eea
while 
\bea
\hbox{Fig.~2b} & = & -g\Tr\left(\Ycal m^2\right)\int d^d k 
\int d^4\theta\, \theta^2\thbar^2 V(\theta,\thbar) 
D^2\frac{\delta^{(4)} (\theta - \theta')}{k^2}\Dbar^2
|_{\theta = \theta'}\nn
& = & -g\Tr\left(\Ycal m^2\right)
\int d^4\theta\, \theta^2\thbar^2 V(\theta,\thbar)\int \frac {d^d k}{k^2}.
\eea
The first term from Eq.~(\ref{oness})\ vanishes via $\Tr\Ycal = 0$, and the 
second may be reduced using the identity Eq.~(\ref{didenta})\
to give
\be
\hbox{Fig.~2a} = g\Tr\left(\Ycal m^2\right)
\int \frac{d^d k}{k^4}\int d^4\theta V(\theta,\thbar)
e^{2\theta\kslash\thbar}.
\label{onessb}
\ee
If we expand the exponential in Eq.~(\ref{onessb}), the 
quadratically divergent $\theta^2\thbar^2$ term cancels Fig.~2b, while 
the remaining term reproduces the component calculation, Eq.~(\ref{compone}).

\section {The two loop calculation}

In this section we discuss the two-loop calculation of $\bxhat$ in  some
detail. Calculations of $\beta$-functions for soft-breaking parameters
may be carried out in components, or using the spurion formalism.
Indeed, as mentioned earlier, in the case of $\beta_h$, $\beta_M$ and
$\beta_{m^2}$  the fact that the spurion  diagrams are only
logarithmically divergent means that these
quantities have simple all-orders expressions in terms of $\gamma$  and
$\beta_g$. However, as we have emphasised, individual diagrams 
contributing to $\beta_{\xi}$ are quadratically divergent. This means
that if, for example,  we represent a $h^{ijk}$ vertex in superspace by
$h^{ijk}\theta^2$, then  we cannot simply factor the $\theta^2$ out,
because it can be  ``hit'' by a superspace $D$-derivative; indeed, as is
clear  from the one-loop calculation, the contribution when the
$\theta^2$  is not ``hit'' will not give a logarithmic divergence, and
must cancel. The simple relationship between  a graph with a $h^{ijk}$
and the corresponding one with  a supersymmetric Yukawa vertex which
holds for the soft breaking $\beta$-functions  is thereby lost.
Nevertheless, the spurion formalism may  still be used. In this section
we shall describe both the spurion approach  and the component
calculation. Normally a superspace calculation would be expected to be
more efficient than the component version. In this case,  however, we
shall see that the advantages of the spurion calculation are by no means
so obvious. The fact that in components the $D$ insertion can only be on
a scalar line considerably reduces the number  of diagrams in this case.

\bigskip
\epsfysize= 2.0in
\centerline{\epsfbox{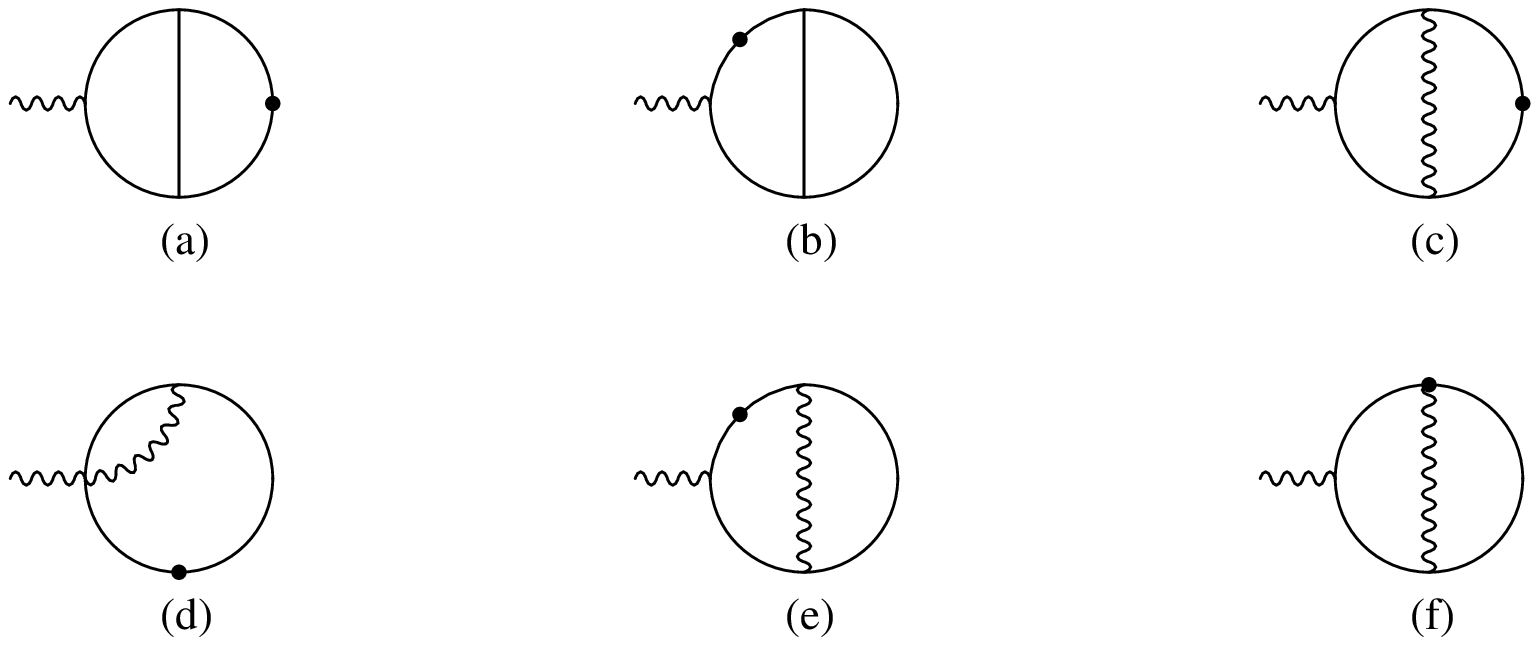}}
\in
{\it \noindent Fig.~3:
Feynman diagrams for the two-loop calculation in superspace. 
Solid lines are chiral propagators and internal wavy lines are vector 
propagators. Blobs denote $m^2$ insertions.}
\bigskip
\out

The two-loop diagrams in the spurion formalism are depicted in Fig.~3. 
Standard superspace manipulations are used to reduce the graphs to basic
momentum  integrals, together with a single remaining $\int d^4\theta$;
by power  counting, the logarithmically divergent contributions come
from terms with no $\theta$s and $\thbar$s remaining in the integrand. 
Some useful identities are collected in Appendix~A. Note that we have
omitted graphs with a mass insertion on the leftmost vertex, where the
external $V$ is attached; these graphs, like Fig.~2b,  do not contribute
to the logarithmic divergence, and are cancelled by  the quadratic 
divergences  (terms with an integrand involving $\theta^2\thbar^2$) 
from the graphs shown.  We have also omitted a graph like Fig.~3d, but
with the mass insertion on the  rightmost vertex, because it also gives
rise to a quadratic divergence only.  The divergent contributions to
$\xi^B$ from each graph are listed in Table~1. 
\vskip3em
\begin{center}
\begin{tabular}{|c|c|c|c|c|c|c|c|}\hline
& a      & b           & c     & d
     & e     & f          \\ \hline
Fig.~3    &$-JS_1$& 0&$8JS_2$
& $-4JS_2$&$4JS_2$&$-4JS_2$\\ \hline
Fig.~5   & $-JS_1$&  $2JS_2$         & $4JS_2$ 
&$-2JS_2$& &              \\ \hline      
\end{tabular}
\end{center}
\in
{\noindent \it Table 1: Results for two-loop Feynman diagrams\hfill}
\out

Here
$J$ denotes the standard two loop momentum integral shown in
Fig.~4, and also
\be
S_1=\Tr\left[\Ycal m^2Y^2\right], \qquad S_2=g^2\Tr\left[\Ycal^3 m^2\right].
\ee 
%\vfil\eject
\bigskip
\epsfysize= 1.0in
\centerline{\epsfbox{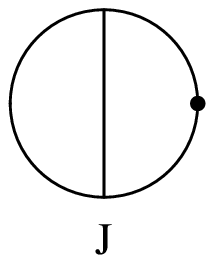}}
\in
{\it \noindent Fig.~4:
Momentum integral for the two-loop calculation. The dot denotes a 
double propagator. }
\bigskip
\out

The calculation of $J$ proceeds as follows (note that here and in all 
subsequent integrals we subtract all subdivergences):
\bea
J & = & \int\,\frac{d^d k d^d q}{q^2 (k-q)^2 (k^2 + m_0^2)^2} - 
\frac{2}{\lf\epsilon}\int\,\frac{d^d k}{(k^2 + m_0^2)^2}\nn
& = & \frac{2}{(4\pi)^d\epsilon^2}
\left(1+\frac{\epsilon}{2} -\gamma\epsilon\right)
-\frac{4}{(4\pi)^{\frac{d}{2}+2}\epsilon^2}\left(1
-\frac{\gamma\epsilon}{2}\right)\nn
& = & \frac{1}{(\lf)^2}\left(-\frac{2}{\epsilon^2}+\frac{1}{\epsilon}\right).
\eea
We have ignored contributions of the form $I^2$, 
where 
\be
I = \int\frac{d^d k}{(k^2 + m_0^2)^2},
\ee
such as that from Fig.~3b, because $I^2$ has no  simple pole after
sub-divergence subtraction; $I^2$ is the simplest possible example of a 
{\it factorised\/} Feynman integral, which quite generally give no
simple  pole\cite{jjold}. Subsequently we will ignore any graph which
reduces  to factorised form.

Thus using the simple pole given by 
\be
J_{\rm simple}
={1\over{\llf\epsilon}},
\ee
and recalling that to get the $L$-loop contribution to the $\beta$-function 
we need to multiply the Feynman diagram simple pole result 
by $L$, we find that at two loops we have 
\be
\lf\bxhat = 2g\Tr\left[\Ycal m^2\right]
-4g\Tr\left[\Ycal m^2 \gamma^{(1)}
\right]+\cdots
\label{exactb}
\ee
so we see that in fact only $A_1$ is non-zero through this order. 

\bigskip
\epsfysize= 2.0in
\centerline{\epsfbox{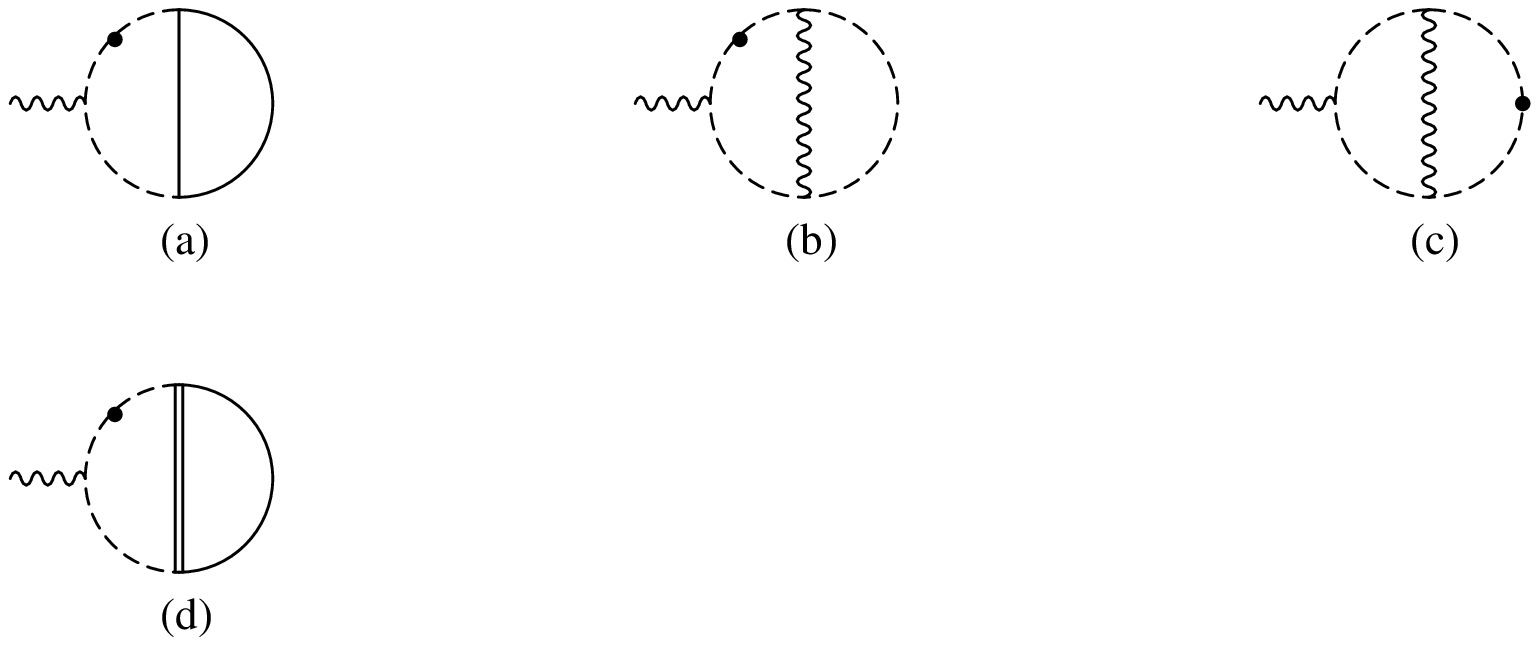}}
\in
{\it \noindent Fig.~5:
Feynman diagrams for the two-loop calculation in components.
Dashed lines are scalar propagators, solid lines are chiral 
fermion propagators, double solid lines are gaugino propagators 
and internal 
wavy lines are vector propagators. Blobs denote $m^2$ insertions.}
\bigskip
\out

The calculation may equally well be performed in the component formalism.
The relevant diagrams are shown in Fig.~5. As we mentioned earlier,
there are relatively few; indeed, fewer than in the spurion case. Their 
divergent contributions are again listed in Table~1, and upon adding we find 
again the result of Eq.~(\ref{exactb}).  It is apparent from Fig.~5 that 
there is no \DRED/\DREDp distinction at this order, because the vector boson 
couples only via the $\phi^*\phi W_{\mu}$ vertex, which projects out the 
$\epsilon$ scalar. 
A further consistency check is provided by Eq.~(\ref{appm});
since
\be
\lf\beta_g = g^3\Tr\left[\Ycal^2\right] 
-2g^3\Tr\left[\Ycal^2\gamma^{(1)}\right]+\cdots
\label{thrb}
\ee
we see that Eq.~(\ref{exactb}) is indeed consistent with Eq.~(\ref{appm}).
Finally, it is easy to verify that our result reproduces the relevant terms 
from the calculation of $\beta_{m^2}$ (with $D$-eliminated) 
presented in Refs.~\cite{jj}, \cite{mv}. (The other two-loop calculation of the
soft $\beta$-functions\cite{yam} was performed with $D$ uneliminated.) 

\section{The three loop results}

We have calculated  
$\bxhat^{(3)\DREDp}$ in full. As we found in the previous section, the 
calculation in terms of component fields is generally more straightforward than 
that using the spurion formalism. In the case of terms 
proportional to $m^2\Ycal Y^4$ we have performed both spurion and 
component calculations, while  for $m^2\Ycal^3Y^2$-type terms we have used 
the spurion formalism, 
which could be streamlined by systematic use of the 
identities in Appendix~A. Both these calculations were sensitive 
to the check provided by Eq.~(\ref{appm}). The rest of the calculation
was done using components. Although the number of diagrams is large, the
amount of algebra involved in each diagram is not great.  

In both component and superfield formalisms, every graph can be reduced 
to a sum of terms consisting of a product of a group theory factor and 
one of a set of logarithmically divergent three loop graphs, which are shown 
in Fig.~6.

\bigskip
\epsfysize= 2.0in
\centerline{\epsfbox{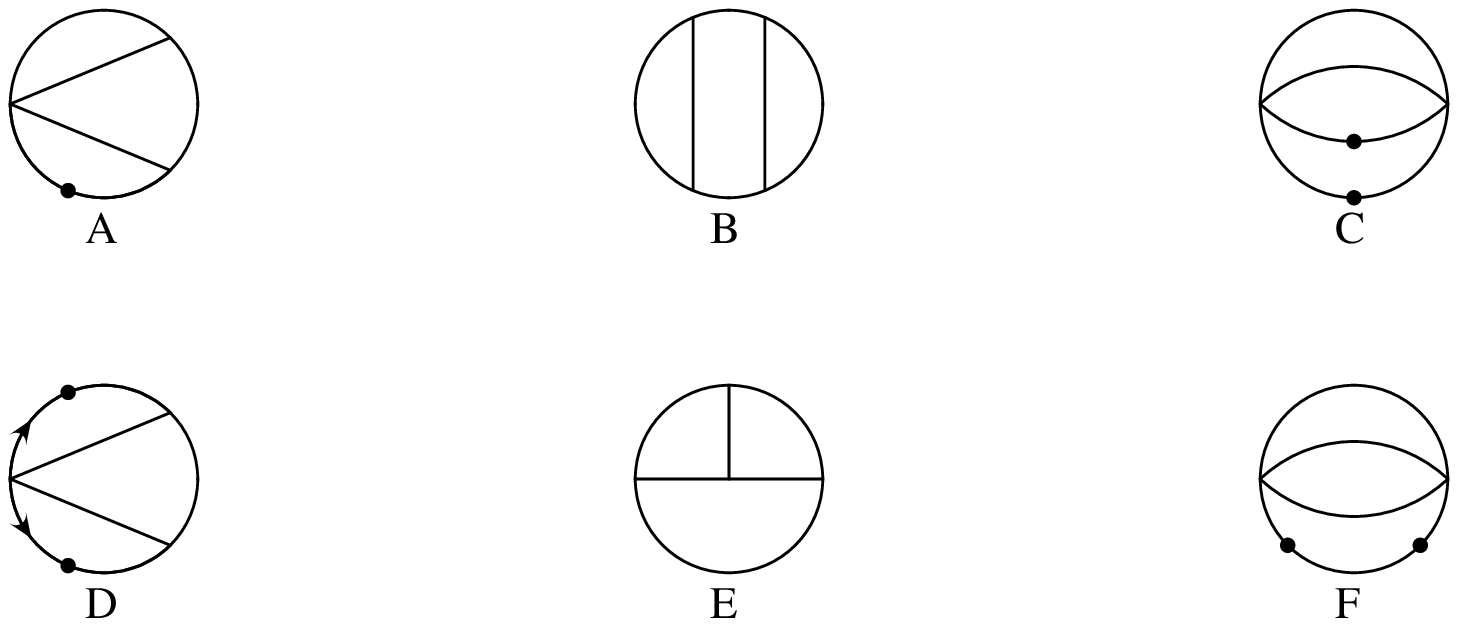}}
\in
{\it \noindent Fig.~6: 
Momentum integrals for the three-loop calculation. One/two dots denote a
double/triple propagator. The two arrows in D denote contracted 
momenta.} 
\bigskip

\out
These graphs may be evaluated by the introduction 
of infra-red regulator masses as described for $J$ in the previous section. 
The results for the 
simple pole contributions (after subtraction of subdivergences) are 
as follows:
\bea
A_{\rm simple} & = & {4\over3}{1\over{\lllf\epsilon}}, \qquad
B_{\rm simple} = -{2\over3}{1\over{\lllf\epsilon}}, \qquad C_{\rm
simple}={2\over3}{1\over{\lllf\epsilon}}, \nn
D_{\rm simple} & = & -{2\over3}{1\over{\lllf\epsilon}}, \qquad
E_{\rm simple}=4\zeta(3){1\over{\lllf\epsilon}}, \qquad
F_{\rm simple}=-{3\over4}{1\over{\lllf\epsilon}}.
\label{integres}
\eea   

We relegate details of the calculation to Appendix~B. The final result 
may be written as follows:
\bea
\lllf\frac{\bxhat^{(3)\DREDp}}{g} &=& -6(\lf)^2\Tr\left[\Ycal m^2
\gamma^{(2)}\right]-4\Tr\left[WP\Ycal\right]- 
\frac{5}{2}\Tr\left[HH^*\Ycal\right]\nn
&+& 2\Tr\left[P^2m^2\Ycal\right]
-24g^2\zeta(3)\Tr\left[W\Ycal^3\right]\nn
&+& 12\zeta(3)g^2\Tr\left[M^*H\Ycal^3 
+ \hbox{c.c.}\right]
-144\zeta(3)g^4MM^*\Tr\left[\Ycal^5\right]\nn
\label{exactc}
\eea
where\cite{jj} 
\be
W^i{}_j =
(\frac{1}{2}Y^2m^2 +\frac{1}{2}m^2Y^2 +h^2)^i{}_j
+2Y^{ipq}Y_{jpr}(m^2)^r{}_q -8g^2MM^*(\Ycal^2)^i{}_j,
\ee

\be
H^i{}_j=h^{ikl}Y_{jkl}+4g^2M(\Ycal^2)^i{}_j
\label{newX}
\ee
and 
\be
\llf\gamma^{(2)i}{}_j =\left[-Y_{jmn}Y^{mpi}-2g^2(\Ycal^2)^p{}_j
\delta^i{}_n\right]P^n{}_p+2g^4\Tr\left[\Ycal^2\right](\Ycal^2)^i{}_j,
\label{gamtwo}
\ee
with
$(Y^2)^i{}_j =  Y^{ikl}Y_{jkl}, (h^2)^i{}_j =  h^{ikl}h_{jkl}.$
We can now check the $m^2$ terms in this result, using 
Eq.~(\ref{appm}). Replacing 
$m^2$ by $g\Ycal$, we obtain
\be
g\lllf\Tr(\Ycal A_1^{(3)}) =  6X_1 + 12X_3+ 2X_4 
-12g^6\Tr\left[\Ycal^2\right]\Tr\left[\Ycal^4\right],
\label{appmn}
\ee
where 
\bea
X_1 &=& g^2Y^{klm}P^n{}_l(\Ycal^2)^p{}_mY_{knp},
\nn
X_3 &=& g^4\Tr \left[P\Ycal^4\right],\nn
X_4 &=& g^2\Tr\left[P^2 \Ycal^2\right],
\label{inv}
\eea 
in precise agreement with the result for $\beta_g^{(3)}$, given in 
\cite{jjna}, which for an abelian theory is 
\be
\lllf\beta_g^{(3)\DRED}= g\left\{3X_1+6X_3+X_4-6g^6\Tr\left[\Ycal^2\right]\Tr
\left[\Ycal^4\right]\right\}.
\label{bthree}
\ee
(Of course for $\beta_g$ there is no distinction between DRED and 
$\DREDp$). 
Note that $\bxhat^{(3)\DRED}$ would only differ from $\bxhat^{(3)\DRED'}$
by the inclusion of terms of the form $g^5\mtilde^2\Tr\left[\Ycal^5\right]$
and $g^3\mtilde^2\Tr\left[P\Ycal^3\right]$, arising from $\epsilon$-scalar 
mass insertions. We have not calculated these explicitly because it is clear 
they can be removed by a redefinition of $m^2$, as follows:
\be
\delta m^2 = - 2\frac{g^2}{\lf}\mtilde^2 \Ycal^2 
+\alpha_1 \left(\frac{g^2}{\lf}\right)^2\mtilde^2 \Ycal^4
+\alpha_2 \frac{g^2}{(\lf)^2}\mtilde^2 P\Ycal^2,
\ee
where the first term was derived in \cite{jjmvy}. It 
would be interesting to verify that the appropriate redefinition also 
renders the three-loop contribution to $\beta_{m^2}$ independent 
of $\mtilde^2$.

Finally, let us compare our result with the form of $\bxhat^{(3)\DREDp}$ 
that we  obtained in Ref.~\cite{jjdterm}\ (note that we did not there 
distinguish \DRED from \DREDp). We see that our result
Eq.~(\ref{exactc})\ indeed confirms the conjectured form given in 
Eq.~(4.10) of Ref.~\cite{jjdterm}, and that the two then undetermined
constants are given by $\nu_1 = 24\zeta (3)$ and $\nu_2 = 0$.

\section{The conformal anomaly trajectory}

The following set of equations provide an exact solution to the 
renormalisation group equations for $M, h$ and $m^2$:
\begin{mathletters}   
\label{result}\bea
M &=& M_0{\beta_g\over g}, \label{result:1}\\
h^{ijk}&=&-M_0\beta_Y^{ijk},\label{result:2}\\
(m^2)^i{}_j &=& \frac{1}{2}|M_0|^2\mu\frac{d\gamma^i{}_j}{d\mu}.
\label{result:3}
\eea
\end{mathletters}
Moreover, these solutions indeed hold if the only source of 
supersymmetry breaking is the conformal anomaly, when $M_0$ is 
in fact the gravitino mass.

This set of soft breakings has generated considerable interest; 
but there are clear difficulties for the MSSM, since  it is
easy to see that sleptons are predicted to have  negative
$\hbox{(mass)}^2$.  Most studies of this scenario have resolved this
dilemma by adding a constant $m_0^2$, presuming another source of
supersymmetry breaking. A non-zero $\xi$ alone is not an alternative,
unfortunately, as is  easily seen from Eq.~(\ref{lsusyb}); the two selectrons,
for example, have  oppositely-signed 
hypercharge so one of them at least remains
with  negative $\hbox{(mass)}^2$. This stumbling block may be overcome
by introducing an extra $U_1$\cite{jjnew}\cite{carena}; for alternative 
treatments see Refs.~\cite{amsbrefs}, \cite{alts}.    

It is immediately obvious that, given Eq.~(\ref{result}), 
there is a RG invariant 
solution for $\xi$ through two loops (for $\bxhat$) given by: 
\be
\lf\xi = g|M_0|^2\Tr\left[\Ycal (\gamma-\gamma^2)\right],
\label{exactxi}
\ee
since differentiating with respect to $\mu$ and using 
Eq.~(\ref{result:3})  leads at once to Eqs.~(\ref{appf}), (\ref{exactb}). 
Interestingly, however, this result for $\xi$ {\it vanishes\/} 
at leading  and next-to-leading order, since one easily demonstrates that 
\be
\Tr\left[\Ycal \gamma^{(1)}\right] = 0
\label{simpa}
\ee
and  
\be
\Tr\left[\Ycal \gamma^{(2)}\right] =    
\Tr\left[\Ycal (\gamma^{(1)})^2\right].
\label{simpb}
\ee
It is interesting to ask 
whether the trajectory can be extended beyond two loops, and whether  it
in fact continues to vanish order by order. We have shown  that there is
indeed a generalisation of Eq.~(\ref{exactxi}) to at least three  
loops (for $\bxhat$), and
that at this order the result for $\xi$ is non-zero.

Our result is as follows: 
\be
\frac{\xi^{\DREDp}}{g|M_0|^2} =
(\lf)^{-4}\left\{-3I_1 
-12\zeta(3)\left(I_2
-2g^6\Tr\left[\Ycal^2\right]\Tr\left[\Ycal^5\right]\right)\right\},
\label{exactxib}
\ee
where
\bea
I_1 &=& \Tr\left[\Ycal P^3\right]
-\frac{1}{2}(\Ycal)^i{}_j Y^{jkl}Y_{imn}P^m{}_k P^n{}_l
+ 2g^2 \Tr\left[\Ycal^3 P^2\right]
-2g^4\Tr\left[\Ycal^2\right]\Tr\left[\Ycal^3 P\right]\nn
I_2&=& g^2(\Ycal^3)^i{}_jY^{jkl}Y_{ikm}P^m{}_l+
g^2\Tr\left[\Ycal^3 P^2\right] + 2g^4\Tr\left[\Ycal^5 P\right].
\label{conjectb}
\eea
It is easy to verify that 
the result of  taking $\mu\frac{\pa}{\pa\mu}$ of Eq.~(\ref{exactxib})
is  identical to that obtained by substituting
Eqs.~(\ref{result}) in 
Eqs.~(\ref{exactb}), (\ref{exactc}). This is a 
non-trivial result
in that  the number of candidate terms  for inclusion in Eq.~(\ref{exactxib})
is  considerably  less than the number of distinct terms which arise
when Eq.~(\ref{exactb}), (\ref{exactc})  are placed on the RG trajectory. We
therefore conjecture that the  trajectory extends  to all orders.  

It is natural to ask what the result for $\bxhat^{(3)}$ is in the NSVZ 
scheme, which is obtained (at the relevant order) by the 
redefinitions\cite{jjf}
\bea
\llf\delta g &=& -\frac{1}{2}g^3\Tr\left[P\Ycal^2\right]\nn
\llf\delta M &=& -Mg^2\left\{\Tr\left[P\Ycal^2\right]
-2g^2\Tr\left[(\Ycal^2)^2\right]\right\}
+\frac{1}{2}g^2 h^{ikl}Y_{jkl}(\Ycal^2)^j{}_i.
\label{drnz}
\eea
It is straightforward to show that in order to obtain the results
Eqs.~(\ref{appf}) and (\ref{appm}) in the NSVZ scheme,  
we must also redefine $\xi$ as follows:
\be
\llf\delta \xi = -\frac{1}{2}g^2\Tr\left[P\Ycal^2\right]\xi
-g \Tr\left[m^2P\Ycal\right].
\label{xirdfn}
\ee
The effect of this is to replace Eq.~(\ref{exactxib}) by 
\be
\frac{\xi^{\NSVZ}}{g|M_0|^2} = (\lf)^{-4}\left\{-4I_1
-12 \zeta(3)\left(I_2 -2g^6\Tr\left[\Ycal^2\right]
\Tr\left[\Ycal^5\right]\right) \right\},
\label{conjectcd}
\ee
and Eq.~(\ref{exactc}) by 
\bea
\lllf\frac{\bxhat^{(3)\NSVZ}}{g} 
&=&-4(\lf)^2\Tr\left[\Ycal m^2 
\gamma^{(2)}\right]
-\frac{5}{2}\left(2\Tr\left[WP\Ycal\right]+
\Tr\left[HH^*\Ycal\right]\right)\nn
&-&24g^2\zeta(3)\Tr\left[W\Ycal^3\right]
+12\zeta(3)g^2\Tr\left[M^*H\Ycal^3 
+ \hbox{c.c.}\right]\nn
&-&144\zeta(3)g^4MM^*\Tr\left[\Ycal^5\right].
\label{exactd}
\eea
It is disappointing that this expression does not immediately suggest 
an all orders result. At this point it is worth recalling that, while 
to connect the \DREDp and \NSVZ schemes via Eq.~(\ref{drnz})\ we
redefined  $g$ and $M$, there exists also a redefinition of $Y$
(involving $\zeta (3)$)  which has the pleasant  property of extending
to three loops the existence of finite $N = 1$ theories\cite{finthree}. 
Unfortunately this redefinition disturbs Eq.~(\ref{Ajy}), which leads
one to imagine that  there might be a combined redefinition of $m^2,Y$
that both preserves Eq.~(\ref{Ajy}) and simplifies $\bxhat^{(3)}$. We
have not yet succeeded in  constructing such a transformation. 

\section{Final remarks}

We have presented a detailed, and we hope a reasonably self-contained 
description of the calculation of $\bxhat$ through three loops. It is
intriguing that in the abelian case we are unable to  express the
renormalisation of the theory completely in terms of  $\beta_g$ and
$\gamma$, which, in the non-abelian case,  suffice to describe the 
renormalisation of both the unbroken theory, and also the theory with
the standard soft terms.  Although there exists perturbatively a
solution related to the AMSB solution for the soft parameters, once
again we are unable at the moment to extend this solution to all orders.

The next step is obviously an extension of our calculation to the case
of a product gauge group including both abelian and non-abelian factors,
such as the MSSM; this is not a trivial deduction from the results we
have presented. Although it is clear that if $\xi$ is assumed to be
small at  gauge unification, then it does not have much effect at low
energies, it  should be remembered that this is an assumption, and that
the MSSM has one more parameter than is commonly supposed.

\acknowledgements
 
This work was supported in part by a Research Fellowship from the
Leverhulme Trust. S.P. was supported by a PPARC Graduate Studentship.
We thank John Gracey for conversations. 

%%%%%%%%%%%%%%%%%%%%%%%APPENDICES
\appendix
\section{$D$-algebra identities}
\label{sec:dalg}
In this Appendix we list some identities that we found useful 
in  superspace calculations of contributions to $\bxhat$. (An early 
reference for superspace calculations incorporating soft breaking is 
Ref.~\cite{neto}.)  
The soft terms given in Eq.~(\ref{softsup})\ are treated as insertions in 
the superfield diagrams and standard superspace manipulations 
may then be used to reduce supergraphs
to ordinary momentum space integrals. It is
convenient to denote the momentum space version of the superspace covariant
derivatives by
\bea
(D_{p1})_{\alpha} & = & 
\frac{1}{2}\left[
\frac{\pa}{\pa\theta_1^{\alpha}}-\pslash_{\alpha\adot}\thbar_1^{\adot}\right],
\nn
(\Dbar_{p1})_{\adot} & = & -\frac{1}{2}\left[ 
\frac{\pa}{\pa\thbar_1^{\adot}}-\pslash_{\alpha\adot}\theta_1^{\alpha}
\right],\nn
\eea   
where
\be
\pslash_{\alpha\adot}\equiv p_{\mu}\sigma^{\mu}_{\alpha\adot}
\equiv i\sigma^{\mu}_{\alpha\adot}\pa_{\mu}.
\ee
We then have the fundamental supersymmetry algebra
\be
\left\{(D_{p1})_{\alpha},(\Dbar_{p1})_{\adot}\right\} = 
\frac{1}{2}\pslash_{\alpha\adot}.
\ee

We find
\be
D_{p1}^2\theta_1^2=-e^{\theta_1\pslash\thbar_1}, \quad
\Dbar_{p1}^2\thbar_1^2=-e^{-\theta_1\pslash\thbar_1},
\ee
and moreover
\bea
D_{p1}^2\Dbar_{p1}^2\theta_1^2\thbar_1^2 D_{p1}^2\Dbar_{p1}^2
\delta_{12} & = & 
e^{\theta_1\pslash\thbar_1 + \theta_2\pslash\thbar_2},\nn
\Dbar_{p1}^2 D_{p1}^2\theta_1^2\thbar_1^2 \Dbar_{p1}^2 D_{p1}^2
\delta_{12} & = &
e^{-(\theta_1\pslash\thbar_1 + \theta_2\pslash\thbar_2)},
\label{didenta}
\eea
where
\be
\delta_{12}=\delta^{(4)}\left(\theta_1 - \theta_2\right) .
\ee
We also have
\begin{mathletters}
\label{didentb}
\bea
\delta_{12}\Dbar_{q1}^2D_{q1}^2e^{2\theta_1\pslash\thbar_1}
\Dbar_{q1}^2 D_{q1}^2\delta_{12}
&=&\delta_{12}\left[(p-q)^2-2p^2\theta_1\qslash\thbar_1
+2q^2\theta_1\pslash\thbar_1+p^2q^2\theta_1^2
\thbar_1^2\right],\label{didentb:1}\\
\delta_{12}D_{q1}^2\Dbar_{q1}^2e^{2\theta_1\pslash\thbar_1}
D_{q1}^2\Dbar_{q1}^2\delta_{12}
&=&\delta_{12}\left[(p+q)^2+2p^2\theta_1\qslash\thbar_1
+2q^2\theta_1\pslash\thbar_1+p^2q^2\theta_1^2
\thbar_1^2\right].\label{didentb:2}
\eea
\end{mathletters}
Finally,
\begin{mathletters}
\label{didentc}
\bea
\delta_{12}D_{r1}^2\Dbar_{r1}^2 [(p-q)^2 &-&2p^2\theta_1\qslash\thbar_1
+2q^2\theta_1\pslash\thbar_1
+p^2q^2\theta_1^2\thbar_1^2]D_{r1}^2\Dbar_{r1}^2\delta_{12}\nn
& = & \delta_{12}\left[(p-q)^2r^2+2q^2p.r-2p^2q.r+p^2q^2\right]
+\hbox{terms in}\quad \theta_1, \thbar_1,\label{didentc:1}\\
\delta_{12}\Dbar_{r1}^2D_{r1}^2 [(p-q)^2 &-&2p^2\theta_1\qslash\thbar_1
+2q^2\theta_1\pslash\thbar_1
+p^2q^2\theta_1^2\thbar_1^2]\Dbar_{r1}^2D_{r1}^2\delta_{12}\nn
&=& \delta_{12}\left[(p-q)^2r^2-2q^2p.r+2p^2q.r+p^2q^2\right]
+\hbox{terms in}\quad \theta_1,\thbar_1.\label{didentc:2}
\eea
\end{mathletters}
Note that the right-hand sides of Eqs.~(\ref{didentb:1}), (\ref{didentb:2}) 
are related by $q\leftrightarrow-q$, and similarly those of 
Eqs.~(\ref{didentc:1}), (\ref{didentc:2}) are related by $r\leftrightarrow-r$.
\section{three loop details}
In this Appendix we give a complete graph-by-graph description of the 
three-loop calculation. We start by giving a list of the distinct tensor
structures involved:
\bea
T_1 & = &(Y^2)^i{}_j Y^{jkl}Y_{ikm}(m^2\Ycal)^m{}_l,\quad
T_2 = (Y^2)^i{}_j Y^{jkl}Y_{imn}(m^2)^m{}_k\Ycal^n{}_l, \quad
T_3 = \Tr\left[Y^2Y^2m^2\Ycal\right],\nn
T_4 & = & g^2\Tr\left[ Y^2 m^2\Ycal^3\right], \quad 
T_5=g^2Y^{ikl}Y_{imn}(m^2\Ycal)^m{}_k (\Ycal^2)^n{}_l,\nn 
T_6 & = & g^2Y^{ikl}Y_{imn}(m^2)^m{}_k(\Ycal^3)^n{}_l,\quad
T_7 = g^4\Tr\left[m^2\Ycal^5\right], \quad T_8=g^4\Tr\left[\Ycal^2\right]
\Tr\left[m^2\Ycal^3\right],\nn
T_9 & = & Y^{ikl}Y_{imn}h_{jkl}h^{pmn}\Ycal^p{}_j,\quad
T_{10} =  \Tr[Y^2h^2\Ycal], \quad T_{11}=g^2\Tr[h^2\Ycal^3],\nn
T_{12} & = & g^2Mh_{ikl}Y^{jkl}(\Ycal^3)^i{}_j, \quad
T_{13} =  g^2MM^*\Tr[Y^2\Ycal^3], \quad T_{14}= g^4MM^*\Tr[\Ycal^5].
\eea
We now give a list of diagrams contributing to these tensor structures,
with the exception of $T_7$, for which there are a very large number of 
separate diagrams; note that when $m^2$ is replaced by $g\Ycal$, $T_7$
and  only $T_7$ produces $\Tr\left[\Ycal^6\right]$, so that we can in fact
infer the coefficient of $T_7$  in our final result via Eq.~(\ref{appm}).
We did, however, perform the explicit $T_7$  calculation, and indeed
obtained the expected result.  
\vfill\eject
\epsfysize= 2.0in
\centerline{\epsfbox{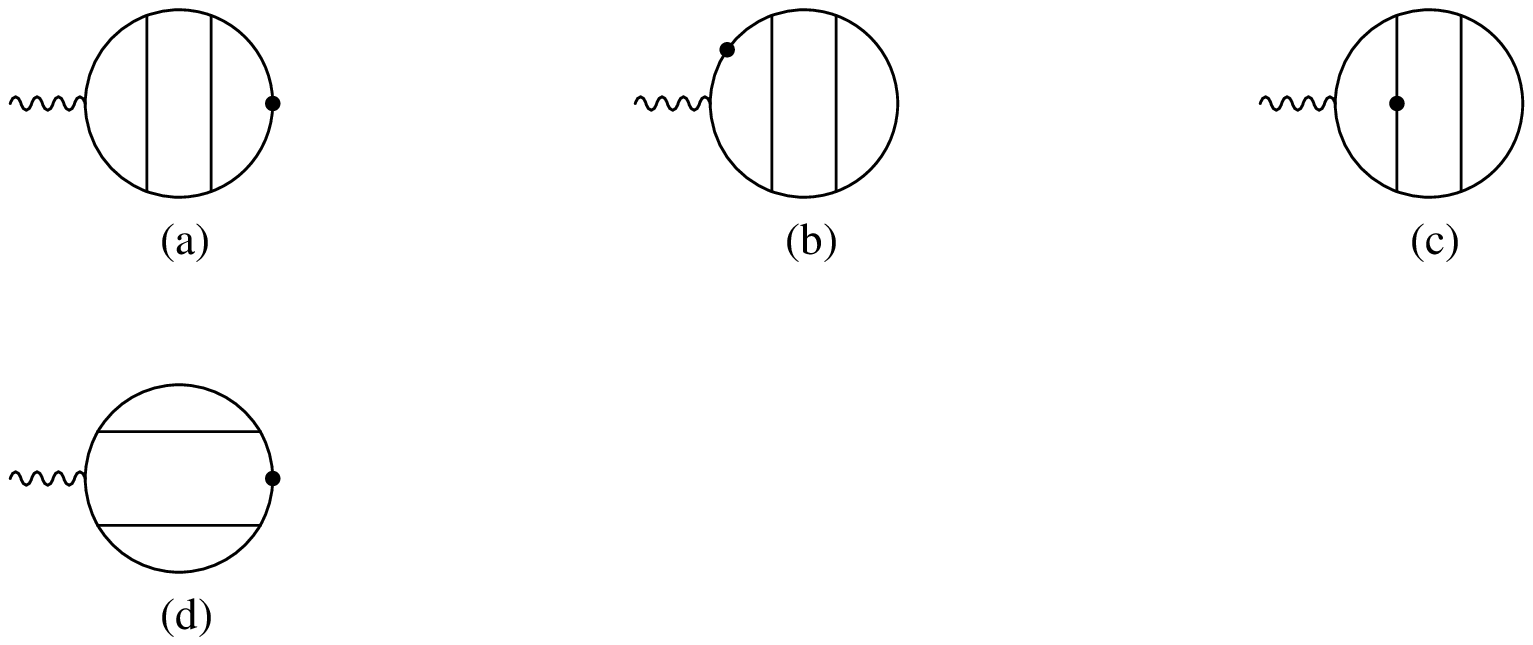}}

\in
{\it \noindent Fig.~7:
Feynman diagrams in superspace for the three-loop contribution of the form
$m^2Y^4\Ycal$, i.e. $T_{1\cdots 3}$. Blobs denote $m^2$ insertions.}
\bigskip
\out

\bigskip
\epsfysize= 2.0in
\centerline{\epsfbox{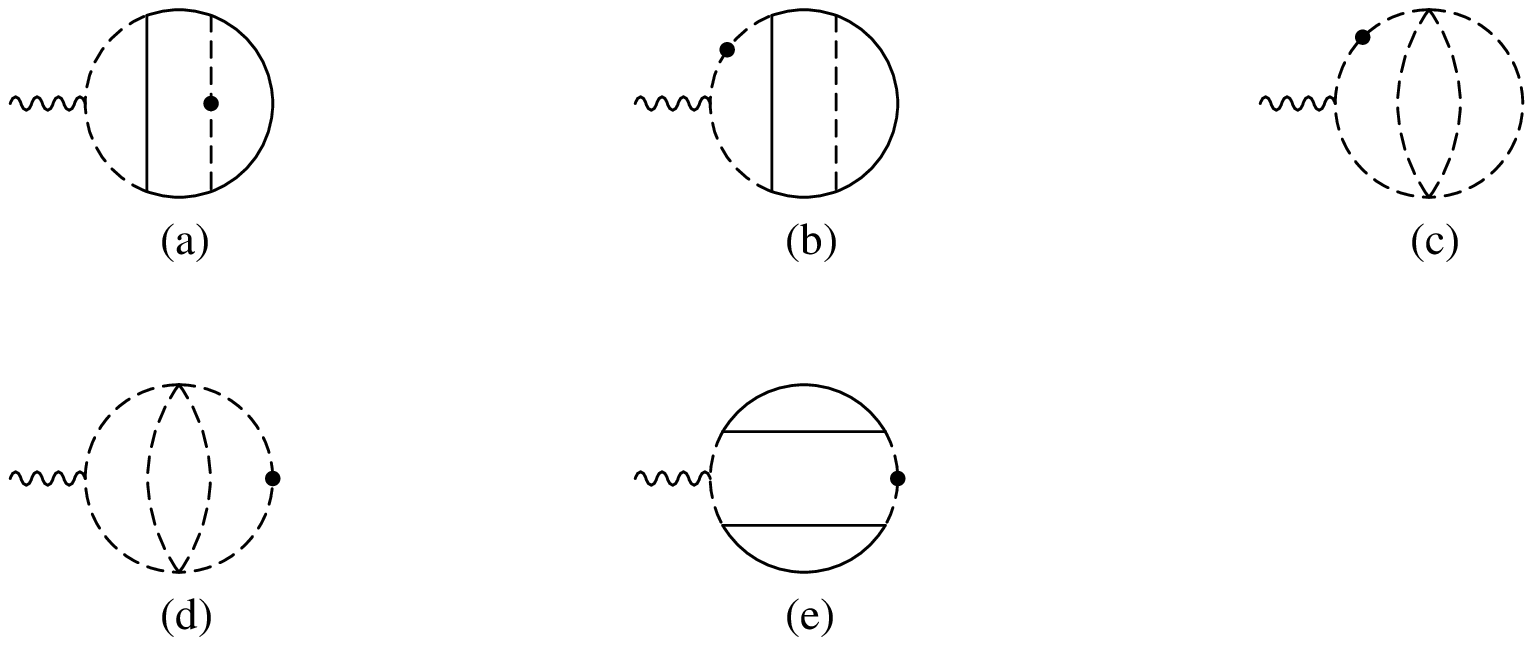}}

\in
{\it \noindent Fig.~8:
Feynman diagrams in components for the three-loop contribution of the form
$m^2Y^4\Ycal$, i.e. $T_{1\cdots 3}$.}
\bigskip
\out

\bigskip
\epsfysize= 6.0in
\centerline{\epsfbox{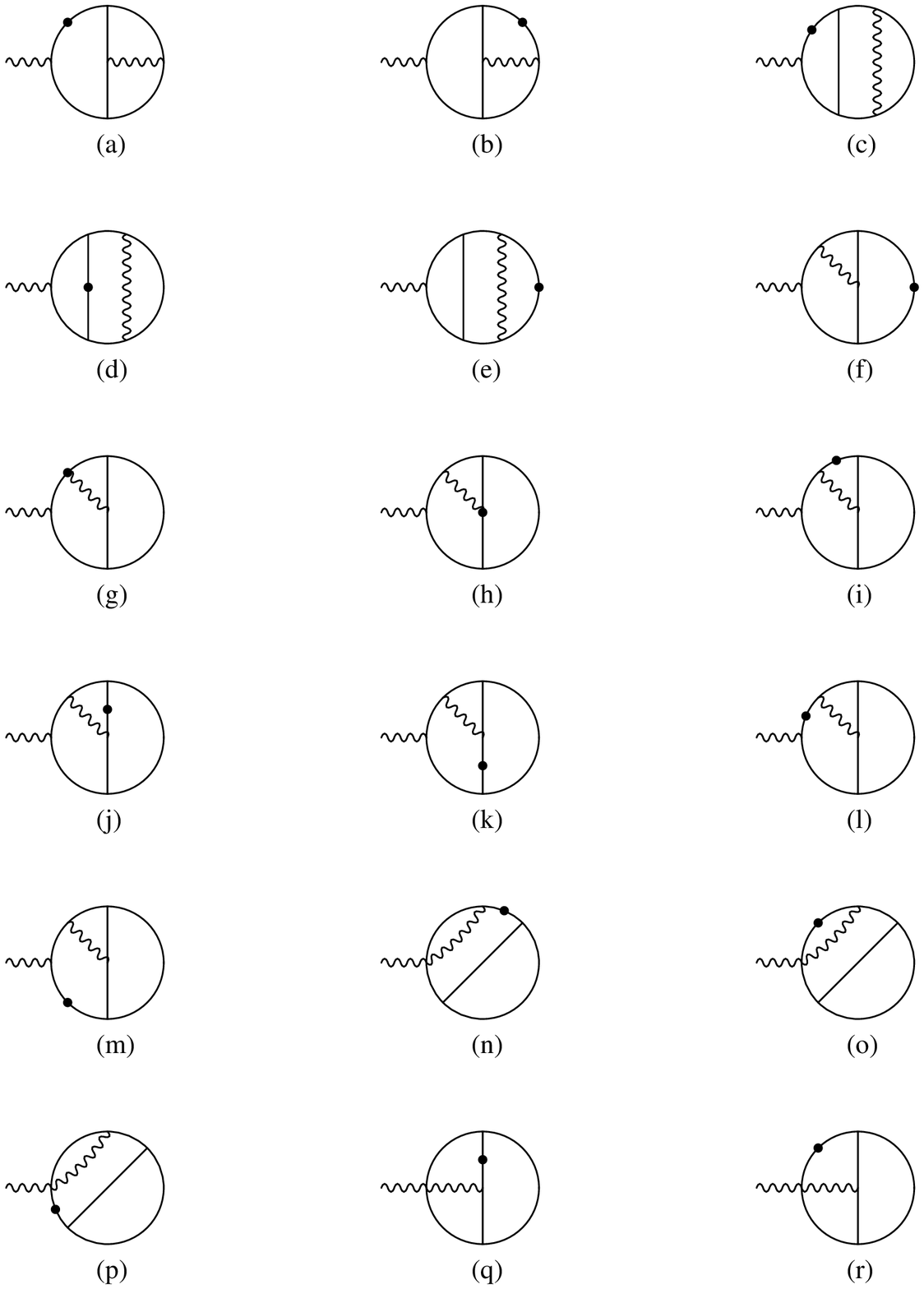}}

\in
{\it \noindent Fig.~9:
Feynman diagrams in superspace for the three-loop contribution of the form 
$g^2m^2Y^2\Ycal^3$, i.e. $T_{4\cdots 6}$.}  
\bigskip
\out

\bigskip
\epsfysize= 4.0in
\centerline{\epsfbox{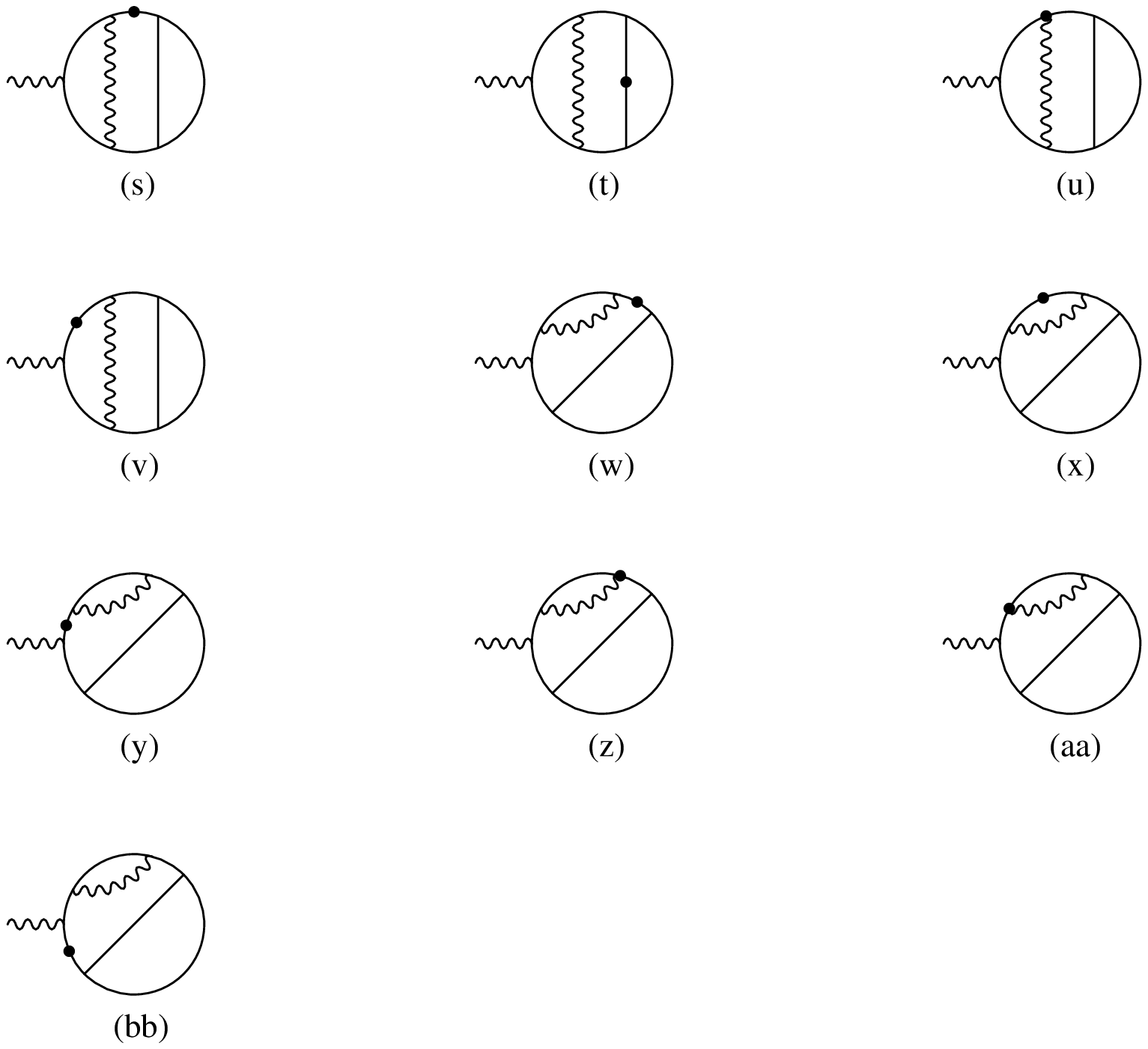}}

\in
{\it \noindent Fig.~9 continued.}
\bigskip
\out

\epsfysize= 3.0in
\centerline{\epsfbox{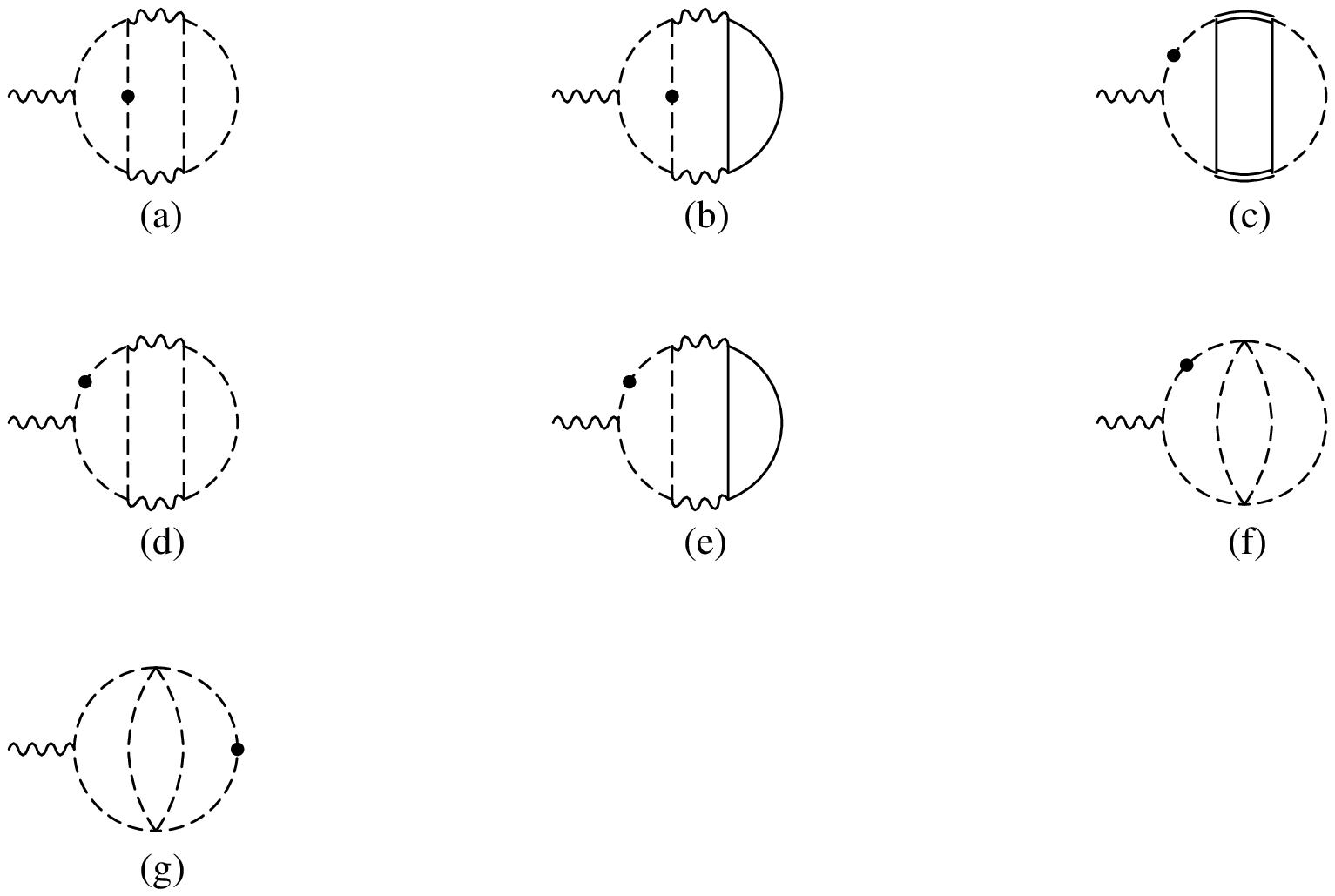}}
\in
{\it \noindent Fig.~10:
Feynman diagrams in components for the three-loop contribution to 
$T_8$.}
\bigskip
\out

\bigskip
\epsfysize= 2.0in
\centerline{\epsfbox{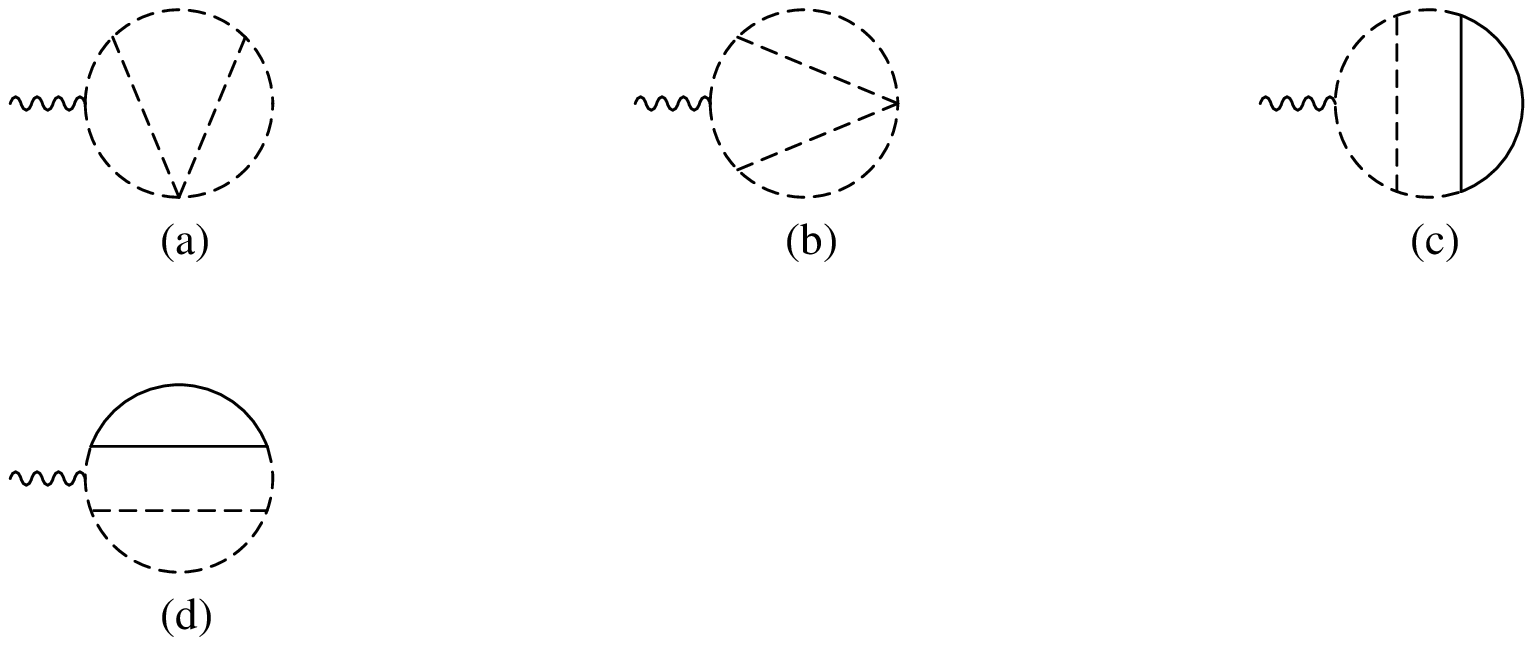}}   
\in
{\it \noindent Fig.~11:
Feynman diagrams in components for the three-loop contribution to
$T_{9}, T_{10}$.}
\bigskip
\out

\bigskip
\epsfysize= 3.0in
\centerline{\epsfbox{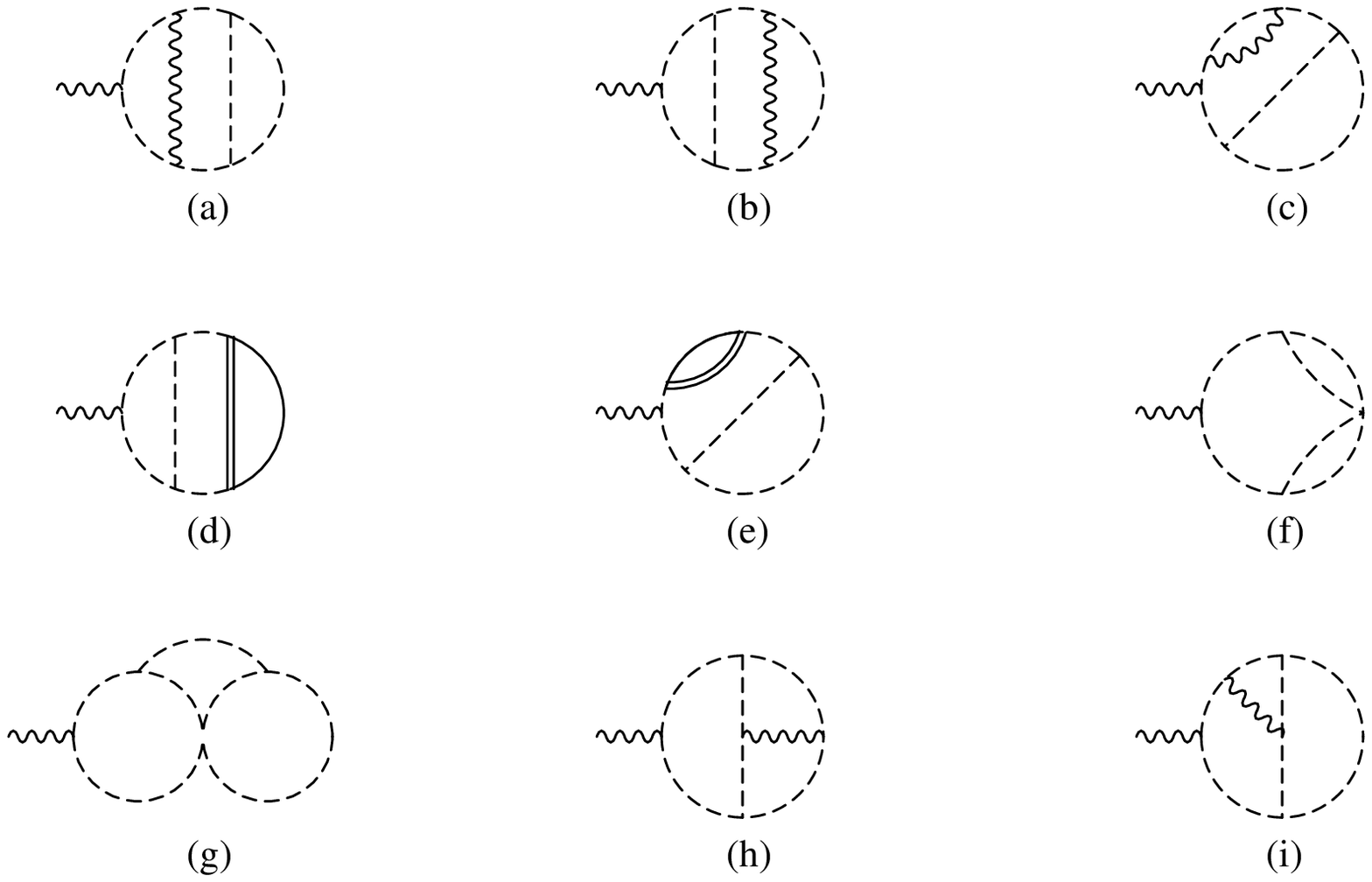}}
\in
{\it \noindent Fig.~12:
Feynman diagrams in components for the three-loop contribution to
$T_{11}$.}
\bigskip
\out

\bigskip
\epsfysize= 1.0in
\centerline{\epsfbox{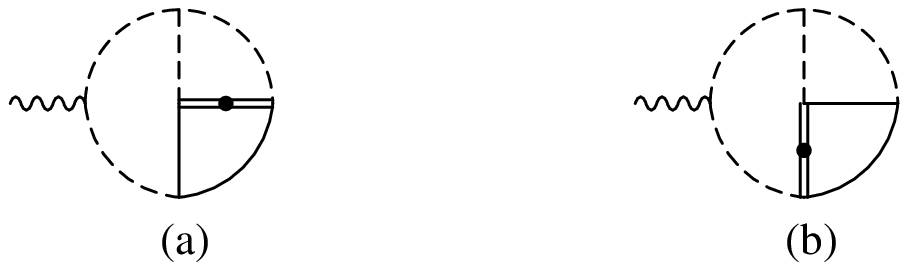}}
\in
{\it \noindent Fig.~13:
Feynman diagrams in components for the three-loop contribution to 
$T_{12}$. Blobs denote gaugino mass insertions.}
\bigskip
\out

\bigskip
\epsfysize= 2.0in
\centerline{\epsfbox{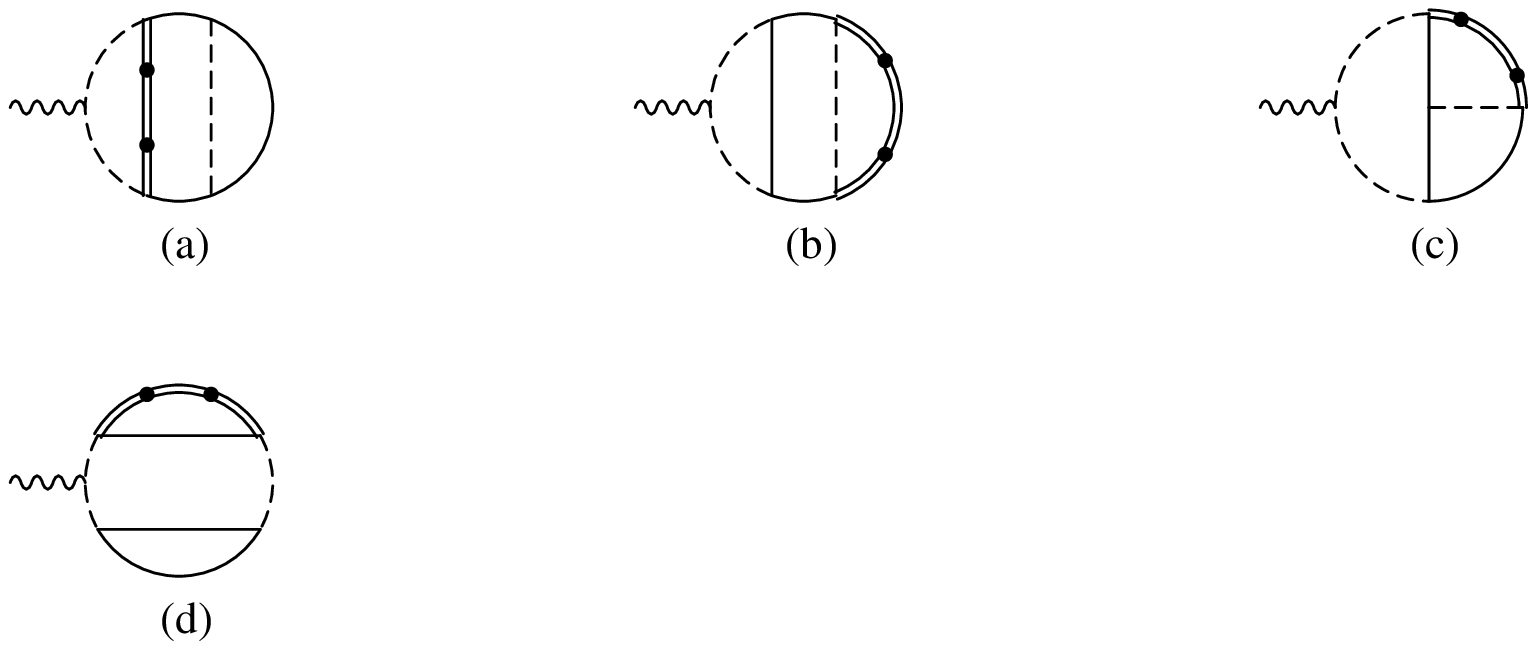}}
\in
{\it \noindent Fig.~14:
Feynman diagrams in components for the three-loop contribution to
$T_{13}$.}
\bigskip
\out

\bigskip
\epsfysize= 3.0in
\centerline{\epsfbox{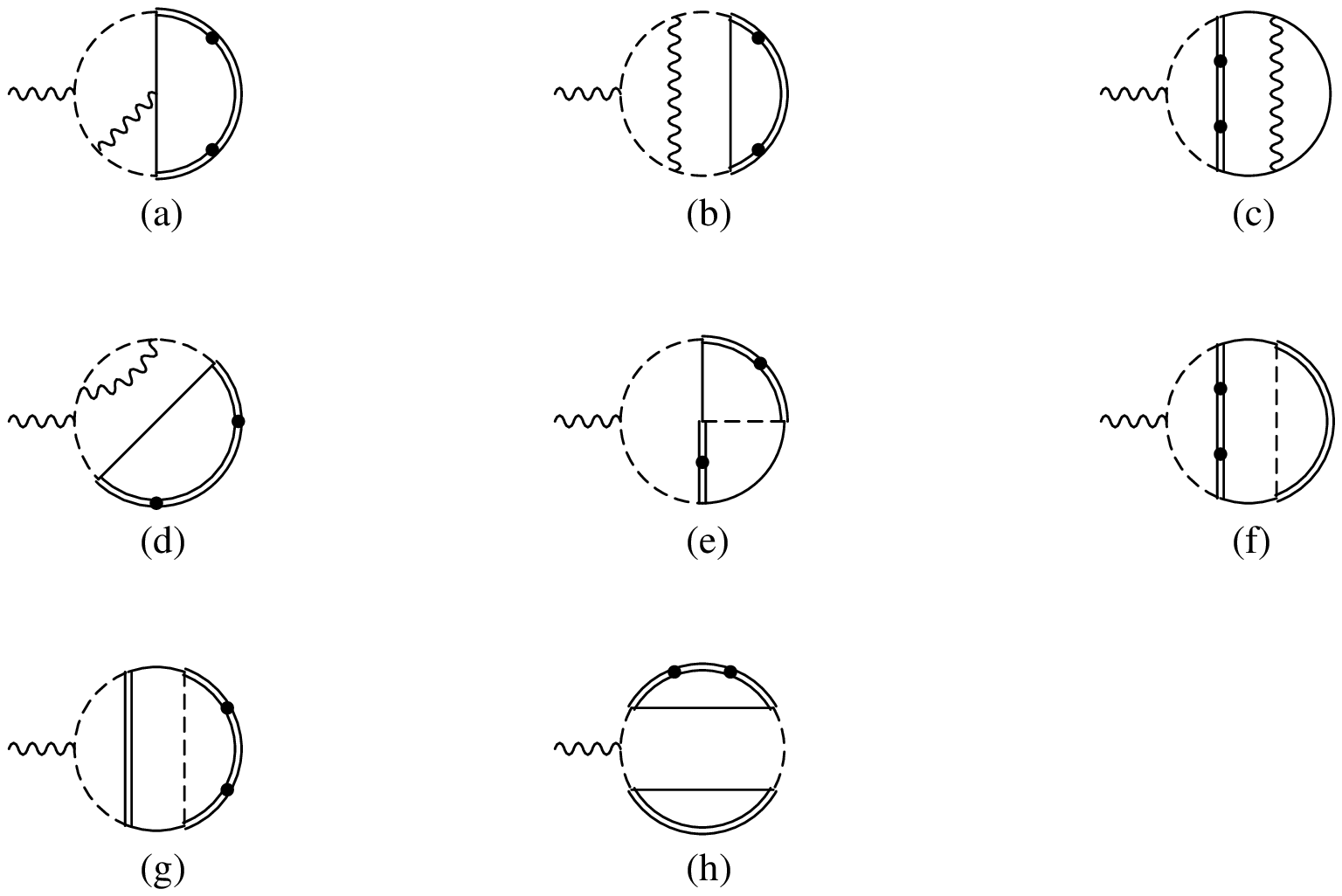}}
\in
{\it \noindent Fig.~15:
Feynman diagrams in components for the three-loop contribution to 
$T_{14}$.}
\bigskip
\out

\bigskip

We begin with a comparison between superspace and component 
formalisms. The results from Fig.~7 (the superspace calculation) are
\be
\hbox{Fig.~7a}=-\frac{1}{2}(B+2D)(T_1+T_2),\quad \hbox{Fig.~7b}=AT_1,\quad
\hbox{Fig.~7c}=\frac{1}{2}CT_2, \quad \hbox{Fig.~7d}=\frac{3}{4}BT_3;
\ee
while from Fig.~8 (the component calculation) we find:
\bea
\hbox{Fig.~8a}=-\frac{1}{2}(B+C+2D)(T_1+T_2), &\qquad& 
\hbox{Fig.~8b}=(A-F)T_1,\nn
\hbox{Fig.~8c}=FT_1, \qquad \hbox{Fig.~8d}&=&\frac{1}{2}C(T_1+2T_2), \quad
\hbox{Fig.~8e}=\frac{3}{4}BT_3.
\eea
(Here and elsewhere, we combine diagrams which  clearly give
identical  results, up to symmetry factors.  Consider for example,
Fig.~8d. Because the scalar fields  are complex, this diagram represents
two distinct (by the usual rules) Feynman  diagrams.) 
The totals of Figs.~7,~8 are manifestly identical, which is
a good check on our spurion rules. 

The results from Figs.~9, 10, 12, 14, 15 are given in
Tables~2--6 respectively:
\medskip
\begin{center}
\begin{tabular}{|c|c|c|}\hline
a& $-2E$&$\frac{1}{2}T_4-T_5$    \\ \hline
b&$-4A$&$\frac{1}{4}T_4+\frac{1}{2}T_5+T_6$\\ \hline
c&$-4A$&$T_5$\\ \hline
d&$-2C$&$-T_5-T_6$\\ \hline
e&$-2(2A+B+2D)$&$-\frac{1}{2}T_4$\\ \hline 
f&$-4A$&$-\frac{1}{4}T_4+ \frac{1}{2}T_5$\\ \hline
g&$4A$&$-\frac{1}{2}T_4$\\ \hline
h&$4A$&$\frac{1}{4}T_4-\frac{1}{2}T_5-T_6$\\ \hline
i&$-4(4A-C-E)$&$-\frac{1}{2}T_4$\\ \hline
j&$-4(2A-2D-E)$&$\frac{1}{4}T_4-\frac{1}{2}T_5-T_6$\\ \hline
k&$-8(A+D)$&$\frac{1}{4}T_4-\frac{1}{2}T_5-T_6$\\ \hline
l&$-4A$&$-\frac{1}{2}T_4$\\ \hline
m&$-4A$&$-\frac{1}{2}T_4$\\ \hline
n&$2B$&$T_4$\\ \hline
o&$2A$&$T_4$\\ \hline
p&$2B$&$T_4$\\ \hline
q&$4A$&$\frac{1}{4}T_4-\frac{1}{2}T_5-T_6$\\ \hline
r&$4A$&$-\frac{1}{2}T_4$\\ \hline
s&$-4(A+B)$&$T_4$\\ \hline
t&$-2(2A+B+2D)$&$T_6$\\ \hline
u&$2A$&$T_4$\\ \hline
v&$-2A$&$T_4$\\ \hline
w&$-2B$&$T_4$\\ \hline
x&$-4(A+B)$&$T_4$\\ \hline
y&$-2B$&$T_4$\\ \hline
z&$2B$&$T_4$\\ \hline
aa&$2B$&$T_4$\\ \hline
bb&$-2B$&$T_4$\\ \hline
\end{tabular}
\end{center}
\in
{\noindent \it Table~2: Results for Fig.~9. \hfill}
\out
\vfill\eject

\begin{center}
\begin{tabular}{|c|c|c|c|c|c|c|c|}\hline
& a + b     & c & d +e     & f        &g\\ \hline
Fig.~10 & $(C-4A-2B)$ &$4(A-F)$&$(2B+2F-4A)$&
$2F$
&$C$\\ \hline
\end{tabular}
\end{center}
\in
{\noindent \it Table~3: Results for Fig.~10 (all multiplied by
$T_8$) \hfill}
\out

The results from Fig.~11 are given by 
\be
\hbox{Fig.~11a}=-\frac{1}{2}AT_9,
\quad\hbox{Fig.~11b}=\frac{1}{4}BT_9,
\quad\hbox{Fig.~11c}=-\frac{1}{4}AT_{10},
\quad\hbox{Fig.~11d}=\frac{1}{2}BT_{10}.
\ee

\medskip
\begin{center}
\begin{tabular}{|c|c|c|c|c|c|c|c|c|c|}\hline
& a      & b           & c     & d
     & e     & f        &g &h&i\\ \hline
Fig.~12   & $-A-B$
& $A$&$-2B$&$-A$&$2B$&$\frac{1}{2}B$
&$-A$&$\frac{1}{2}(4A-B-2E)$&$A-B-E$\\ \hline
\end{tabular}
\end{center}
\in
{\noindent \it Table~4: Results for Fig.~12 (all multiplied by
$T_{11}$)\hfill}
\out

\medskip
The results from Fig.~13 are:
\be
\hbox{Fig.~13a}=(E-2A)(T_{12}+(T_{12})^*),\quad
\hbox{Fig.~13b}=B(T_{12}+(T_{12})^*).
\ee
\medskip
\begin{center}
\begin{tabular}{|c|c|c|c|c|c|c|c|}\hline
& a      & b           & c     & d \\ \hline
Fig.~14   & $-C$
& $-(A+B+2D)$&$2C$&$2(A-B)$\\ \hline
\end{tabular}
\end{center}
\in
{\noindent \it Table 5: Results for Fig.~14 (all multiplied by
$T_{13}$)\hfill}
\out
\medskip
\begin{center}
\begin{tabular}{|c|c|c|c|c|c|c|}\hline
& a      & b           & c     & d
     & e     & f        \\ \hline
Fig.~15    & $8(E+\frac{1}{2}C-A+\frac{1}{2}B)$
& $4(B-C)$&$-2C$&$8(B-A)$&$8(\frac{1}{2}B+\frac{1}{2}E - A)$
&$-2C$\\ \hline\hline
&g&h&&&&\\ \hline
Fig.~15&$4(A+B+2D)$&$8(A-B)$&&&&\\ \hline 
\end{tabular}
\end{center}
\in
{\noindent \it Table~6: Results for Fig.~15 (all multiplied by 
$T_{14}$)\hfill}
\out

The final total is obtained by combining the tables, 
substituting the simple pole results 
for $A,B\cdots F$ from Eq.~(\ref{integres}), and 
multiplying by $3$ (for 3 loops):
\bea
\lllf\frac{\bxhat^{(3)\DREDp}}{g}&=&7T_1+4T_2-\frak{3}{2}T_3
+\left[10-24\zeta(3)\right]T_4 -12T_5\nn
&+&16\left[1-3\zeta(3)\right]T_6-16T_7-12T_8 -\frak{5}{2}T_9\nn
&-&2T_{10}
+8\left[1-3\zeta(3)\right]T_{11}-2\left[5-6\zeta(3)\right]
\left(T_{12}+T_{12}^*\right)\nn
&+&16T_{13}-8\left[13-18\zeta(3)\right]T_{14},
\eea
which can easily be recast into the form given in Eq.~(\ref{exactc}). 
(As indicated earlier, we have suppressed details of the $T_7$ computation). 

\end{document}